\documentclass[sigconf]{acmart}
\usepackage[ruled,vlined]{algorithm2e}
\usepackage{subfigure}
\usepackage{array}
\usepackage{graphicx}
\usepackage{multirow}
\usepackage{bm}
\usepackage{enumitem}
\usepackage{setspace}
\usepackage{tcolorbox}
\newcommand{\tabincell}[2]{\begin{tabular}{@{}#1@{}}#2\end{tabular}}
\AtBeginDocument{%
  \providecommand\BibTeX{{%
    \normalfont B\kern-0.5em{\scshape i\kern-0.25em b}\kern-0.8em\TeX}}}

\author{Shangbin Feng}
\affiliation{%
  \institution{Xi'an Jiaotong University}
  \city{Xi'an}
  \country{China}}
\email{wind\_binteng@stu.xjtu.edu.cn}

\author{Herun Wan}
\affiliation{%
  \institution{Xi'an Jiaotong University}
  \city{Xi'an}
  \country{China}}
\email{wanherun@stu.xjtu.edu.cn}

\author{Ningnan Wang}
\affiliation{%
  \institution{Xi'an Jiaotong University}
  \city{Xi'an}
  \country{China}}
\email{mrwangyou@stu.xjtu.edu.cn}

\author{Jundong Li}
\affiliation{
  \institution{University of Virginia}
  \city{Charlottesville}
  \country{USA}}
\email{jundong@virginia.edu}

\author{Minnan Luo}
\affiliation{
  \institution{Xi'an Jiaotong University}
  \city{Xi'an}
  \country{China}}
\email{minnluo@xjtu.edu.cn}

\begin{document}
\fancyhead{}

\settopmatter{printacmref=false} % Removes citation information below abstract
\renewcommand\footnotetextcopyrightpermission[1]{} % removes footnote with conference information in first column
\pagestyle{plain} % removes running headers

\title{SATAR: A Self-supervised Approach to Twitter Account Representation Learning and its Application in Bot Detection}


% \author{Shangbin Feng}
% \affiliation{%
%   \institution{Xi'an Jiaotong University}
%   \city{Dalaran}
%   \country{Eastern Kingdoms}}
% \email{wind_binteng@stu.xjtu.edu.cn}

% \author{Herun Wan}
% \affiliation{%
%   \institution{Xi'an Jiaotong University}
%   \city{Stormwind}
%   \country{Eastern Kingdoms}}
% \email{wanherun@stu.xjtu.edu.cn}

% \author{Ningnan Wang}
% \affiliation{%
%   \institution{Xi'an Jiaotong University}
%   \city{Kul' Tiras}
%   \country{Eastern Kingdoms}}`
% \email{mrwangyou@stu.xjtu.edu.cn}

% \author{Minnan Luo}
% \affiliation{
%   \institution{Xi'an Jiaotong University}
%   \city{Dalaran}
%   \country{Eastern Kingdoms}  
% }
% \email{minnluo@xjtu.edu.cn}

\begin{abstract}
  Twitter has become a major social media platform since its launching in 2006, while complaints about bot accounts have increased recently. Although extensive research efforts have been made, the state-of-the-art bot detection methods fall short of generalizability and adaptability. Specifically, previous bot detectors leverage only a small fraction of user information and are often trained on datasets that only cover few types of bots. As a result, they fail to generalize to real-world scenarios on the Twittersphere where different types of bots co-exist. Additionally, bots in Twitter are constantly evolving to evade detection. Previous efforts, although effective once in their context, fail to adapt to new generations of Twitter bots. To address the two challenges of Twitter bot detection, we propose SATAR, a self-supervised representation learning framework of Twitter users, and apply it to the task of bot detection. In particular, SATAR generalizes by jointly leveraging the semantics, property and neighborhood information of a specific user. Meanwhile, SATAR adapts by pre-training on a massive number of self-supervised users and fine-tuning on detailed bot detection scenarios. Extensive experiments demonstrate that SATAR outperforms competitive baselines on different bot detection datasets of varying information completeness and collection time. SATAR is also proved to generalize in real-world scenarios and adapt to evolving generations of social media bots.
\end{abstract}

% <concept_id>10002951.10003260.10003282.10003292</concept_id>
% <concept_desc>Information systems~Social networks</concept_desc>
% <concept_id>10010147.10010257.10010293.10010294</concept_id>
% <concept_desc>Computing methodologies~Neural networks</concept_desc>
% <concept_id>10010147.10010178.10010179</concept_id>
% <concept_desc>Computing methodologies~Natural language processing</concept_desc>
% <concept_significance>300</concept_significance>

% \ccsdesc[500]{Information systems~Social networks}
% \ccsdesc[500]{Computing methodologies~Neural networks}
% \ccsdesc[300]{Computing methodologies~Natural language processing}

% \keywords{Twitter Bot Detection; Social Media; Self-supervised Learning; Representation Learning}

\maketitle
%\vspace{-6pt}
\section{Introduction}
\label{sec:introduction}
Twitter is a popular online social media platform which was released in 2006. Individuals can sign up for a Twitter account to view and publish content of their interests. As reported by Statista\footnote{\url{https://www.statista.com/}}, the number of daily active Twitter users in the United States is over 35 million in the second quarter of 2020\footnote{\url{https://www.statista.com/statistics/970911/monetizable-daily-active-twitter-users-in-the-united-states/}}. Twitter has become not only an essential social platform in people's daily life but also an information publishing venue. The open nature and widespread popularity of Twitter have made itself an ideal target of exploitation from automated programs, also known as bots. These bot accounts are often operated to achieve malicious goals. Bots have been actively involved in many important events, including the elections in the United States and Europe~\cite{10.1145/3308560.3316486, DBLP:journals/corr/Ferrara17aa}. Bots are also responsible for spreading fake news and propagating extreme ideology~\cite{berger2015isis}. These malicious bots try to hide their automated nature by imitating the behaviors of normal users. Across the whole Twittersphere, it is reported that bot accounts for 9\% to 15\% of total active users~\cite{yardi2010detecting}. Since bots jeopardize user experience in Twitter and may even induce undesirable social effects, many research efforts have been devoted to Twitter bot detection. 

The first work to detect automated accounts in social media dates back to 2010~\cite{yardi2010detecting}. Early studies conducted feature engineering and adopted traditional classification algorithms. Three categories of features were considered: (1) user property features~\cite{d2015real}; (2) features derived from tweets~\cite{miller2014twitter}; and (3) features extracted from neighborhood information~\cite{yang2013empirical}. Later, researchers began to propose neural network based bot detection frameworks. Wei \textit{et al.} ~\cite{wei2019twitter} adopted long short-term memory to extract semantics information from tweets. Kudugunta \textit{et al.}~\cite{kudugunta2018deep} proposed a method that combined feature engineering and neural network models. Heuristic methods for bot detection were also put forward recently. Minnich \textit{et al.}~\cite{minnich2017botwalk} proposed a bot detection method based on anomaly detection. Cresci \textit{et al.}~\cite{cresci2016dna} encoded tweets into a string to find out the difference between human and bots in tweeting behaviors.

Despite early successes, ever-shifting social media brought two new challenges to the task of bot detection: generalization and adaptation. The challenge of generalization in social media bot detection demands bot detectors to simultaneously identify bots that attack in many different ways and exploit diversified features on Twitter. Cresci \textit{et al.}~\cite{cresci2017paradigm} points out that Twitter bots attack in different ways such as retweet frauds, malicious hashtag promotion and URL spamming. They also imitate the tweeting behaviour of different types of genuine users, fill out profile items differently and follow each other to boost their follower count. Since Twitter bots are indeed becoming more diversified, a robust Twitter bot detector should therefore address the challenge of generalization to induce real-world impact. However, previous bot detection methods fail to generalize since they only leverage limited user information and are trained on datasets with few types of bots.

Apart from that, the challenge of adaptation in bot detection demands bot detectors to maintain desirable performance in different times and catch up with rapid bot evolution. Cresci \textit{et al.}~\cite{10.1145/3409116}'s investigation shows that bots in the past used to be simple and easily identified, possessing too little profile and friend information to be genuine. However, more recently evolved bots have large numbers of friends and followers, use stolen profile pictures and intersperse malicious tweets with neutral ones. These newly evolved bots often evade existing detection measures, thus a robust bot detector should address the challenge of adaptation to put an end to the arms race between bot evolution and bot detection research. However, previous bot detection measures rely heavily on feature engineering and are not designed to adapt to emerging trends in bot evolution.

% For generalizability, social media bots pursue distinctive goals. Bots in politics aim to spread fake news and incite emotions while phishing bots trick users into malicious URLs. The real-world Twittersphere is home to a diversified motley of bots, where previous methods trained on limited public datasets fall short of generalization. For adaptability, early efforts of bot detection inspired countermeasures by bot operators. Bots are constantly evolving to evade detection, and previous methods fail to adapt to the bot evolution. That being said, there is an urgent need for a bot detector that can generalize on real-world social media and adapt to the evolution of bots.

%However, early success of bot detection inspired countermeasures by bot operators. New generations of bots are harder to identify and they often feature advanced characteristics. Previous methods cannot adapt to bot evolution. Additionally, social media bots often have different targets. Bot users in politics aim to incite strong emotions through fake news and doctored images while phishing bots try their best to trick users into malicious URLs. The real-world Twittersphere is home to a diversified motley of bots, where previous frameworks trained on limited public datasets fall short of generalization. That being said, there is an urgent need for a bot detection method that can (1) adapt to the evolution of bot accounts and (2) generalize on real-world social media platforms.

In light of the two challenges of Twitter bot detection, we propose a novel framework SATAR (\textbf{S}elf-supervised \textbf{A}pproach to \textbf{T}witter \textbf{A}ccount \textbf{R}epresentation learning). SATAR adopts self-supervised learning to obtain user representation and identify bots on social media. Specifically, SATAR jointly encodes tweet, property and neighborhood information of users without feature engineering to promote bot detection generalization. SATAR follows a pre-training and fine-tuning learning schema to adapt to different generations of bots. Our main contributions are summarized as follows:

\begin{itemize} [topsep=4pt, leftmargin=*]
    \item We propose a novel framework SATAR to conduct generalizable and adaptable Twitter bot detection. SATAR is an end-to-end framework that jointly uses semantic, property and neighborhood information of users without feature engineering.
    \item To the best of our knowledge, this paper is the first work to introduce self-supervised representation learning to improve the performance of bot detection.
    % \item SATAR adopts self-supervised representation learning to improve its generalizability and adaptability. It can recognize different types of bots.
    % \item We collect and annotate an up-to-date dataset {\verb|TwiBot-20|}\footnote{The dataset is available at \url{https://github.com/GabrielHam/SATAR}} which is the first comprehensive sample of the Twittersphere. We publicize {\verb|TwiBot-20|} hoping to facilitate further research.
    \item We conduct extensive experiments on three real-world datasets to evaluate SATAR and competitive baselines. SATAR outperforms baselines on all three datasets and is proved to generalize and adapt through further exploration.
\end{itemize}

\noindent In the following, we first review related work in Section ~\ref{sec:relatedwork} and define the task of Twitter bot detection in Section ~\ref{sec:problemdefinition}. Next, we propose SATAR in Section ~\ref{sec:SATAR}, following with extensive experiments in Section ~\ref{sec:experiments}. Finally, we conclude the whole paper in Section ~\ref{sec:conclusion}.

\section{Related Work}
\label{sec:relatedwork}
In this section, we briefly review the related literature on self-supervised learning and Twitter bot detection.
% \subsection{Representation Learning}
% % https://kopernio.com/viewer?doi=10.1145%2F3292500.3330948&token=WzI1MjQ1MjUsIjEwLjExNDUvMzI5MjUwMC4zMzMwOTQ4Il0.bHymzT8vkaVb1-t-UPUG1Z6St9M
% % word-embedding
% % Basic, ELMO, GPT, BERT (archive)
% Representation learning aims to learn a low-dimensional vector to represent high-dimensional information. 
% In natural language processing, the idea of distributed representation of words, also known as word embedding, is extensively studied~\cite{mikolov2013distributed, peters2018deep, radford2018improving, devlin2018bert}. Jiang \textit{et al.}~\cite{10.1145/3308558.3313707} proposed a framework to learn document representations from multiple
% abstraction levels of the document structure. 
% In graph representation learning, neural networks ~\cite{cao2016deep, hinton2006reducing, narayanan2017graph2vec} are widely used to learn the representations. Recently, Peng \textit{et al.}~\cite{peng2020graph} proposed a new concept of graphical mutual information and it outperforms state-of-the-art unsupervised counterpart.
% % Cao \textit{et al.}~\cite{cao2016deep} utilized deep neural networks to capture information conveyed by the graph and argue the advantage of it. 
% % Peng \textit{et al.}~\cite{peng2020graph} proposed a new concept of graphical mutual information and it outperforms state-of-the-art unsupervised counterpart.

\subsection{Twitter Bot Detection}
Traditional bot detection methods mainly focused on extracting basic features from user information. Among them, Gao \textit{et al.}~\cite{gao2012towards} used text shingling and incremental clustering to merge spam messages into campaigns for real-time classification. 
% Cresci \textit{et al.}~\cite{cresci2016dna} applied standard DNA analysis techniques to generate a DNA sequence from user tweeting behaviors. 
Lee \textit{et al.} \cite{lee2013warningbird} proposed to use the redirection of URLs in tweets and Thomas \textit{et al.} \cite{thomas2011design} focused on classification of mentioned websites . 
% Gao \textit{et al.}~\cite{lee2013warningbird} classified tweets without links by analyzing the textual patterns of each tweet.
Other features are also adopted such as information on the user profile ~\cite{lee2011seven}, social networks ~\cite{minnich2017botwalk} and timeline of accounts ~\cite{cresci2016dna}. Yang \textit{et al.}~\cite{yang2013empirical} designed several new features to counter the evolution of modern Twitter bots. Cresci \textit{et al.}~\cite{cresci2018reaction} proposed that confrontation between bot detectors and bot operators is a never-ending arms race. It is also argued that we should refrain from methods that rely on posterior observations.

Neural networks are also adopted to detect Twitter bots because of their strong learning capability. 
Wei \textit{et al.}~\cite{wei2019twitter} employed recurrent neural networks to efficiently capture features across tweets. Kudugunta \textit{et al.}~\cite{kudugunta2018deep} divided user features into account-level features, such as follower count, and tweet-level features, such as the number of hashtags. Both kinds of features and semantic information are used to set up an LSTM-based bot detection framework. Stanton \textit{et al.}~\cite{stanton2019gans} utilized generative adversarial network for spam detection to avoid annotation costs and inaccuracies. 
% Chavoshi \textit{et al.}~\cite{chavoshi2016debot} developed a warped correlation filter to identify correlated user accounts. Minnich \textit{et al.}~\cite{minnich2017botwalk} constructed a bot bank from a small number of seed bots.
% https://dl.acm.org/doi/10.1145/3308560.3316504 & desktop
Alhosseini \textit{et al.}~\cite{ali2019detect} proposed a model based on graph convolutional networks for spam bot detection to leverage both node features and neighborhood information. However, these supervised methods rely heavily on annotated data while relevant datasets are typically limited in size. We draw from self-supervised learning to leverage large quantities of unlabeled data.

\subsection{Self-Supervised Learning}

% The basic idea is to learn representations of the data or to automatically label a dataset. 
In order to use unsupervised dataset in a supervised manner, self-supervised learning frames a special learning task, predicting a subset of entities' information using the rest.
% Self-supervision tasks for deep learning generally involve training the network with filling in the hiding parts of a complex information.
% 1: https://arxiv.org/pdf/1708.07860.pdf
% 2: https://arxiv.org/pdf/1809.07207.pdf
% 3: Original BERT
% 4: https://openreview.net/forum?id=YYvtmM9TPw
% 5: https://www.ri.cmu.edu/publications/improving-robot-navigation-through-self-supervised-online-learning/
% 6: https://arxiv.org/abs/1905.06566
As a promising learning paradigm, self-supervised learning has drawn massive attention for its fantastic data efficiency and generalization ability, with many state-of-the-art models following this paradigm \cite{liu2020self}. Doersch \textit{et al.}~\cite{doersch2017multi} combined several self-supervised tasks to jointly train a network. 
Zhai \textit{et al.}~\cite{zhai2019s4l} proposed that semi-supervised learning can benefit from self-supervised learning. 

Self-supervised learning has been used in different domains, such as 
% representation learning ~\cite{doersch2017multi, appalaraju2020towards}, 
% robotics ~\cite{nava2019learning, sofman2006improving}, 
natural language processing ~\cite{devlin2018bert, zhang2019hibert}, 
computer vision ~\cite{oord2016conditional, larsson2016learning} and graph analysis ~\cite{grover2016node2vec, kipf2016variational}.
In natural language processing, self-supervised tasks are designed based on following words ~\cite{radford2018improving} or the whole sentence ~\cite{mikolov2013distributed}. Masked language models are also adopted to  better attend to the content in general ~\cite{devlin2018bert}. In computer vision, adjacent pixels \cite{oord2016conditional, van2016pixel} and the full image \cite{dinh2014nice, dinh2016density} are used for pretext tasks similarly. In graph analysis, self-supervised tasks are designed based on edge attributes ~\cite{dai2018adversarial, tang2015line} or node attributes ~\cite{ding2018semi}.

\section{Problem Definition}
\label{sec:problemdefinition}
Let $U$ be a Twitter user, consisting of three aspects of user information: semantic $T$, property $P$ and neighborhood $N$. Let $T = \{t_i\}_{i=1}^{M}$ be a user's semantic information of $M$ tweets. Each tweet $t_i = \{w_1^i, \cdot \cdot \cdot, w_{Q_i}^i\}$ contains $Q_i$ words. Let $P = \{p_i\}_{i=1}^{R}$ be a user's property information with a total of $R$ properties. Each property $p_i$ could be numerical such as follower count or categorical such as whether the user is verified. Let $N = \{N^f, N^t\}$, where $N^f = \{N_1^f, \cdot \cdot \cdot, N_u^f\}$ are $u$ followings of the user and $N^t = \{N_1^t, \cdot \cdot \cdot, N_v^t\}$ are $v$ followers. Similar to previous research~\cite{yang2020scalable, kudugunta2018deep}, we treat Twitter bot detection as a binary classification problem, where each user could either be human ($y = 0$) or bot ($y = 1$). Formally, we can define the Twitter bot detection task as follows:

\hspace{2pt}

\begin{tcolorbox}[
    standard jigsaw,
    opacityback=0,
    boxrule=0.5pt% this works only in combination with the key "standard jigsaw"
]
\textbf{Problem: Twitter Bot Detection} Given a Twitter user $U$ and its information $T$, $P$ and $N$, learn a bot detection function $f:f(U(T,P,N)) \rightarrow \hat{y}$, such that $\hat{y}$ approximates ground truth $y$ to maximize prediction accuracy.
\end{tcolorbox}

\section{SATAR methodology}
\label{sec:SATAR}
In this section, we present the details of the proposed Twitter user representation learning framework named as SATAR (\textbf{S}elf-supervised \textbf{A}pproach to \textbf{T}witter \textbf{A}ccount \textbf{R}epresentation learning).

% In Section~\ref{subsec:SATARover}, we provide an overview of the proposed framework SATAR. In Section ~\ref{subsec:SATARTSSN} - ~\ref{subsec:SATARCIA}, we formally define the architecture of SATAR and details regarding its four major components. In Section ~\ref{subsec:SATARSSLO}, we provide details about the self-supervised learning schema and present the overall SATAR training algorithm.

\subsection{Overview}
\label{subsec:SATARover}
\begin{figure}
  \centering
  \includegraphics[width=.95\linewidth]{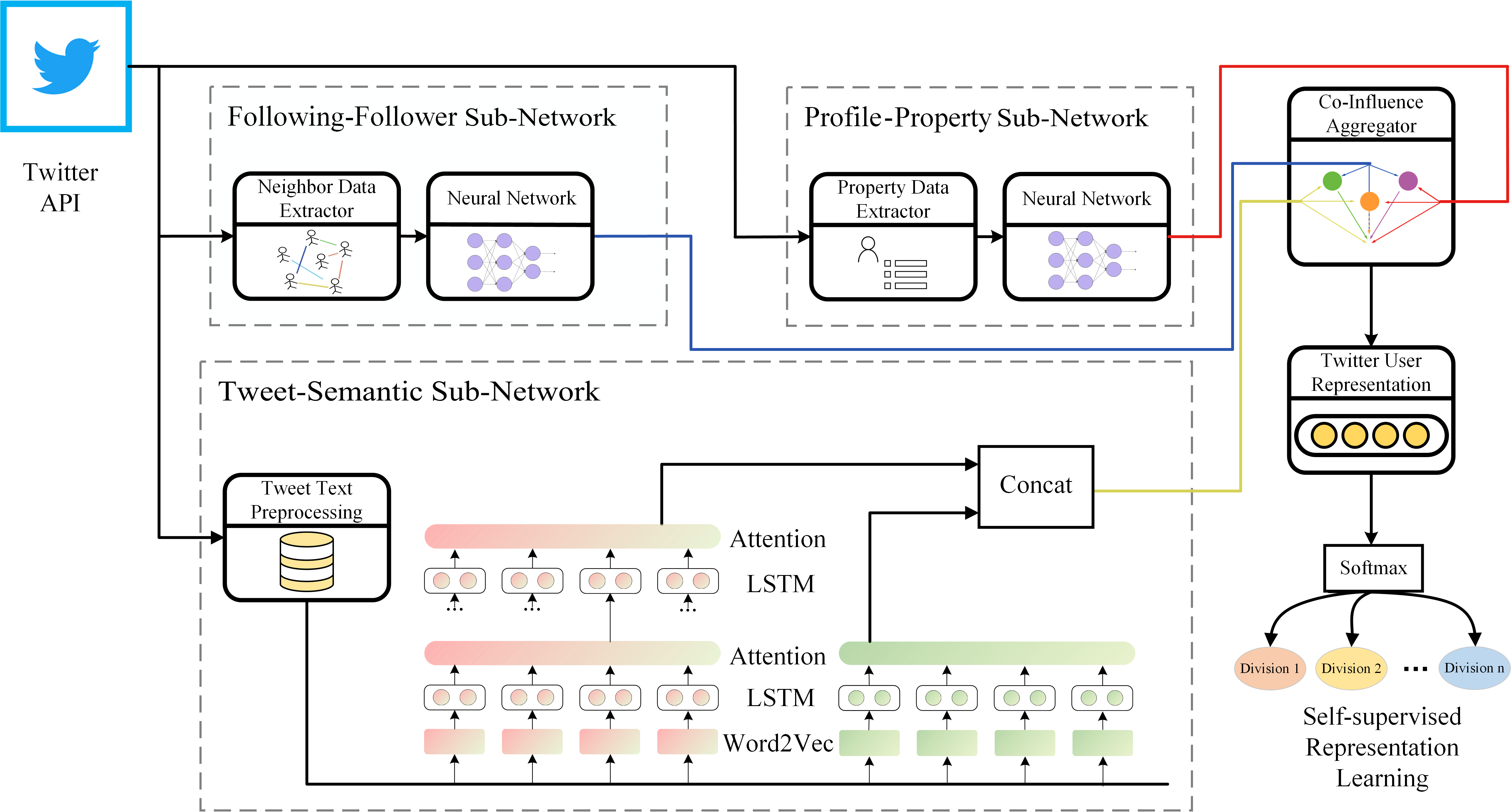}
  \caption{Overview of our proposed framework SATAR.}
  \Description{SATAR architecture in a nutshell}
  \label{fig:SATAR}
\end{figure}

Figure ~\ref{fig:SATAR} illustrates the proposed framework SATAR. It consists of four major components: (1) a tweet-semantic sub-network, (2) a profile-property sub-network, (3) a following-follower sub-network and (4) a Co-Influence aggregator. Specifically, we use the Twitter API\footnote{\url{https://developer.twitter.com/en/products/twitter-api/early-access}} to obtain relevant data regarding a user's semantic, property and neighborhood information. The tweet-semantic sub-network encodes a Twitter user's textual information into $r_s$ with hierarchical RNNs of different depth accompanied by the attention mechanism. The profile-property sub-network encodes a Twitter user's profile properties into $r_p$ with property data encoding and fully connected layers. The following-follower sub-network encodes a Twitter user's neighborhood relationships into $r_n$ with neighborhood information extractor and fully connected layers. Finally, a non-linear Co-Influence aggregator takes the correlation between three aforementioned components into account, generating a representation vector that fully embodies the social status of a specific Twitter user. A softmax layer is then applied for user classification and enables model learning.

\subsection{Tweet-Semantic Sub-Network}
\label{subsec:SATARTSSN}
%Most of the previous works about Twitter bot detection have utilized users' tweet content information. Firstly, hand-picked keywords and feature engineering are pervasive in bot detection endeavors. These approaches extracted information that is conceived as helpful to bot detection, such as URL count ~\cite{AHMED20131120}, hashtag count~\cite{yang2013empirical} and the frequency of spam words~\cite{8508495}. Perceived as effective to begin with, these approaches are generally abandoned due to the inevitable bias introduced in the feature engineering process. With the advent of deep learning, techniques in natural language processing are adopted to capture the semantic information in a specific user's tweets, which shows promising results for bot detection. These efforts treat each tweet as a distinct entity, which is deemed independent from other tweets in evaluating a Twitter user. However, two characteristics of tweeting behaviors would weaken such an assumption of independence. Firstly, Twitter currently has a 280-character limit for tweets, which forces longer texts to become a thread while tweets in a thread often has a coherent meaning. Secondly, a specific user's tweets represent a sequential flow of the user's engagements on social media, but the temporal dependence of different tweets are not considered by existing works.

In this paper, we exploit user semantic information at two different levels, tweet-level and word-level, to capture the tweet content of users. Specifically, words in a user's tweets could be fitted into two hierarchical structures. For tweet-level characterization, as defined in Section~\ref{sec:problemdefinition}, $w_i^j$ denotes the $i$-th word in the $j$-th tweet of the user timeline, and $t_j$ represents the $j$-th tweet of a specific user. We also concatenate temporally adjacent tweets: $\{w_1, \cdot \cdot \cdot, w_K\} = \{w_1^1,  \cdot \cdot \cdot, w_{Q_1}^1, w_1^2, \cdot \cdot \cdot, w^M_{Q_M}\}$, where the total word count $K = \sum_{i=1}^M Q_i$. Thus for word-level characterization, $w_k$ denotes the $k$-th word in the user's tweet history with temporally adjacent tweets concatenated to form a sequence. It is noteworthy that the underlying words are identical between tweet-level and word-level, but their annotations differ according to the user's tweeting behaviors. To jointly leverage user tweet information on these two different levels, we propose tweet-level and word-level encoders of hierarchical RNNs to model tweet text sequences respectively and derive an overall semantic representation for Twitter users. The overall semantic representation for Twitter users are concatenated with results of tweet-level and word-level:

\begin{equation}
    \label{send}
    r_s = concatenation(r^t_s; r^w_s).
\end{equation}

where $r_s^t$ and $r_s^w$ are representations of tweets on tweet-level and word-level.

% \textbf{Tweet Text Pre-Processing}
% To tackle the diversity and irregularity of modern social-media rhetoric, we perform several transformations to the original textual information:
% \begin{itemize}
% \item {Hashtag}: Hashtags in a tweet are converted into $\_tag\_$.
% \item {Mention}: Mentions in a tweet are converted into $\_atsb\_$.
% \item {URL}: URLs in a tweet are converted into $\_link\_$.
% \item {Emoji}: Emotes in a tweet are converted to their textual counterparts using the {\verb|emoji|} python package\footnote{https://pypi.org/project/emoji/}.
% \end{itemize}

% We adopt the aforementioned transformations of tweet text to preserve the universal meaning embodied by $\#$, $@$, emojis and URLs. Please refer to section 8.2 in the appendix for an enumeration of the adopted text pre-processing procedure.

% After cleaning tweet text, we train word embeddings with every tweet in the dataset combined using Word2Vec~\cite{mikolov2013distributed}, producing a vector for each word that represents its meaning in the Twitter context.

\noindent \textbf{Tweet-Level Encoder.}
The tweet-level encoder follows a bottom-up approach. For the $j$-th tweet of a specific user, we first embed words in it with an embedding layer:

\begin{equation}
  \label{sbegin}
  x_i^j = emb(w_i^j), 1 \leqslant i \leqslant Q_j, 1 \leqslant j \leqslant M,
\end{equation}

\noindent where $Q_j$ is the length of the $j$-th tweet, and we use Word2Vec~\cite{mikolov2013distributed} as the embedding layer $emb(\cdot)$. To encode the tweet, a bidirectional RNN processes the tweet in a forward pass and a backward pass. For the forward pass, a sequence of forward hidden states is generated for the $j$-th tweet:
\begin{equation}
    \overrightarrow{h}^t_j = \bigg[\overrightarrow{h}^t_{j,1}, \overrightarrow{h}^t_{j,2}, \cdot \cdot \cdot, \overrightarrow{h}^t_{j,Q_j}\bigg],
\end{equation}

\noindent where the hidden representation for each step is generated by

\begin{equation}
    \overrightarrow{h}^t_{j,i} = RNN\bigg(\overrightarrow{h}^t_{j,i-1}, x_i^j\bigg).
\end{equation}

Here we use LSTM~\cite{lstm} as $RNN(\cdot)$, which is widely adopted to model long-term dependencies in a sequence. For the backward pass, a sequence of backward hidden states is generated similarly:

\begin{equation}
    \overleftarrow{h}^t_j = \bigg[\overleftarrow{h}^t_{j,1}, \overleftarrow{h}^t_{j,2}, \cdot \cdot \cdot, \overleftarrow{h}^t_{j,Q_j}\bigg].
\end{equation}

We concatenate the forward and backward results to form a sequence of word representations in the $j$-th tweet:
\begin{equation}
    h^t_j = \bigg[ h^t_{j,1}, h^t_{j,2},\cdot \cdot \cdot, h^t_{j,Q_j} \bigg],
\end{equation}
% \noindent where the hidden representation for each step is generated by
% \begin{equation}
%     \overleftarrow{h^t_{j,i}} = RNN\bigg(\overleftarrow{h^t_{j,i+1}}, x^t_{j,i}\bigg)
% \end{equation}

\noindent where $h^t_{j,i}=\bigg[ \overrightarrow{h}^t_{j,i}; \overleftarrow{h}^t_{j,i} \bigg]$. Since words in a tweet vary in their contribution to the tweet's overall semantic meaning, the attention mechanism is adopted to aggregate word hidden representations into a tweet vector. Specifically,
\begin{equation}
    \alpha^t_{j,i} = \frac{exp(u^t_{j,i} \cdot v_l^t)}{\sum_{i'}exp(u^t_{j,i'} \cdot v_l^t)},
\end{equation}

\noindent where $u^t_{j,i} = tanh(W_l^t h^t_{j,i}+b_l^t)$ transforms vectors for each word and $v_l^t$, $W_l^t$ and $b_l^t$ are learnable parameters. $\alpha^t_{j,i}$ represents the weight of the $i$-th word in the $j$-th tweet. Finally, the representation of the $j$-th tweet can be obtained as follows:

\begin{equation}
    v^t_j = \sum_i \alpha^t_{j,i}h^t_{j,i}.
\end{equation}

After deriving a vector for each tweet, the tweet-level encoder applies RNN similarly to tweet representations $\{v^t_j\}_{j=1}^M$, generating a forward and a backward sequence. We concatenate the forward and backward results to form a sequence of tweet representations:

% The forward pass generates a sequence of hidden tweet vectors:
% \begin{equation}
%     \overrightarrow{h}^t = \bigg[\overrightarrow{h}^t_1, \overrightarrow{h}^t_2, \cdot \cdot \cdot, \overrightarrow{h}^t_m \bigg]
% \end{equation}

% \noindent where $\overrightarrow{h}^t_i = RNN(\overrightarrow{h}^t_{i-1}, v^t_i)$ and $m$ is the total tweet count of a specific user. For the backward pass, a sequence of hidden tweet representations is similarly generated:
% \begin{equation}
%     \overleftarrow{h}^t = \bigg[\overleftarrow{h}^t_1, \overleftarrow{h}^t_2, \cdot \cdot \cdot, \overleftarrow{h}^t_m \bigg]
% \end{equation}

% \noindent where $\overleftarrow{h^t_i} = RNN(\overleftarrow{h^t_{i+1}}, v^t_i)$. 

\begin{equation}
    h^t = \bigg[ h^t_1, h^t_2, \cdot \cdot \cdot, h^t_M \bigg],
\end{equation}

\noindent where $h^t_i = \bigg[\overrightarrow{h}^t_i;\overleftarrow{h}^t_i\bigg]$. An attention layer is applied to model the influence each tweet has on the overall semantics of the user:
\begin{equation}
    \alpha^t_i = \frac{exp(u^t_i \cdot v^t_h)}{\sum_{i'}exp(u^t_{i'} \cdot v^t_h)},
\end{equation}

\noindent where $u^t_i = tanh(W^t_h h^t_j + b^t_h)$ transforms vectors for each tweet and $v^t_h$, $W^t_h$ and $b^t_h$ are learnable parameters. $\alpha^t_i$ represents the weight of the $i$-th tweet. Finally, the representation of a user's tweet semantics from a tweet-oriented perspective can be obtained as follows:

\begin{equation}
    r^t_s = \sum_i \alpha^t_i h^t_i.
\end{equation}

% \begin{figure*}[h]
%   \centering
%   \includegraphics[width=\linewidth]{semantic_hierarchy_1.png}
%   \caption{Tweet-Level (left) and Word-Level (right) Tweet Text Hierarchy}
%   \Description{Tweet Hierarchy in a nutshell}
%   \label{fig:text_hierarchy}
% \end{figure*}

\noindent \textbf{Word-Level Encoder.}
The word-level encoder concatenates temporally adjacent tweets into a long sequence of words. For the $i$-th word of the sequence, we first embed it with the embedding layer identical to the tweet-level encoder:
\begin{equation}
    x_i = emb(w_i), 1 \leqslant i \leqslant K,
\end{equation}

\noindent where $K$ is the total word count in the temporally concatenated tweets. A bidirectional RNN with attention is adopted to encode the concatenated sequence. For the forward pass, we have: 
\begin{equation}
    \overrightarrow{h}^w = \bigg[ \overrightarrow{h}^w_1,\overrightarrow{h}^w_2,\cdot \cdot \cdot, \overrightarrow{h}^w_K \bigg],
\end{equation}

\noindent where $\overrightarrow{h}^w_i = RNN(\overrightarrow{h}^w_{i-1}, x_i)$ and LSTM is adopted for $RNN(\cdot)$ regarding its particular length. For the backward pass, we have:
\begin{equation}
    \overleftarrow{h}^w = \bigg[ \overleftarrow{h}^w_1,\overleftarrow{h}^w_2,\cdot \cdot \cdot, \overleftarrow{h}^w_K \bigg],
\end{equation}

\noindent where $\overleftarrow{h}^w_i = RNN(\overleftarrow{h}^w_{i+1}, x_i)$. Then we concatenate the forward and backward results to form a sequence of word representations in the user's tweet history:
\begin{equation}
    h^w = \bigg[h^w_1,h^w_2,\cdot \cdot \cdot, h^w_K \bigg],
\end{equation}

\noindent where $h^w_i = \bigg[\overrightarrow{h}^w_i; \overleftarrow{h}^w_i \bigg]$. Then the attention mechanism is applied:
\begin{equation}
    \alpha^w_i = \frac{exp(u^w_i \cdot v^w)}{\sum_{i'}exp(u^w_{i'} \cdot v^w)},
\end{equation}
\noindent where $u^w_i = tanh(W^w h^w_i + b^w)$, $v^w$, $W^w$ and $b^w$ are learnable parameters, $\alpha^w_i$ represents the weight of the $i$-th word in the concatenated sequence. Finally, the representation of a user's tweet semantics from a word-oriented perspective is as follows:
\begin{equation}
    r_s^w = \sum_i \alpha^w_i h^w_i.
\end{equation}

% \vspace{-2pt}
% \noindent \textbf{Overall Semantic Representation.}
% The tweet-semantic sub-network produces an overall representation $r_s$ based on the two encoders:
% \begin{equation}
%     \label{send}
%     r_s = concatenation(r^t_s; r^w_s).
% \end{equation}

\subsection{Profile-Property Sub-Network}
\label{subsec:SATARPPSN}

To avoid the undesirable bias incorporated in feature engineering, the profile-property sub-network utilizes profile properties that could be directly retrieved from the Twitter API. Different encoding strategies are adopted for different types of property data:
\begin{itemize}[leftmargin=*]
    \item There are 15 true-or-false property items in total. We use 1 for true and 0 for false. e.g. “profile uses background image”.
    \item There are 5 numerical property items in total. We apply z-score normalization to numerical properties over the whole dataset. e.g. “favorites count”.
    \item There is one special property item: “location”. We divide locations geographically into different countries and apply one-hot encoding.
\end{itemize}
% \begin{itemize}
% \item {True-or-False Property}: 1 denotes True, 0 denotes False. e.g. whether a user uses the default avatar.
% \item {Real-Valued Property}: z-score normalization is applied to boost numerical robustness while preserving original information:
% \begin{equation}
%     x' = \frac{x - mean(X)}{std(X)}
% \end{equation}
% \noindent where $mean(X)$ is the average of real-value random variable X and $std(X)$ is the standard deviation of X. e.g. user following count.
% \item {Categorical Property}: one-hot encoding is adopted, while the details might vary between different property items. e.g. user location.
% \end{itemize}

It is noteworthy that the follower count of a specific user would not be included in the property vector, which would be part of the self-supervised learning schema presented in Section ~\ref{subsec:SATARSSLO}.

The encoded property items are concatenated to form a raw property vector $u_p$, which is then transformed to produce the Twitter user's property representation $r_p$:
\begin{equation}
    \label{p}
    r_p = ReLU(FC_p(u_p)),
\end{equation}

\noindent where $FC_p(\cdot)$ is a fully connected layer and $ReLU(\cdot)$ is a nonlinearity adopted as the activation function.

\subsection{Following-Follower Sub-Network}
\label{subsec:SATARFFSN}
For user followings, according to Twitter mechanism, their tweets will appear in the timeline and the following behaviors often demonstrate interest in their tweet content. Thus we propose $u_n^{f}$ to model the following relationships:
\begin{equation}
    \label{nbegin}
    u_n^{f} = \frac{1}{\sum_{u\in N^f}TF(u)}\sum_{u\in N^f} TF(u) r_s(u),
\end{equation}

\noindent where $N^f$ denotes the following set of a Twitter user, $TF(u)$ denotes the tweet frequency of user $u$ and $r_s(u)$ is the semantic representation of user $u$ generated by the tweet-semantic sub-network. Tweet frequency $TF$ is approximated by a user's total tweet count divided by account active time, which is the time period between a user's registration and its last update.  Note that $\frac{TF(u)}{\sum_{u'\in N^f} TF(u')}$ represents the proportion that user $u$ appears in one's timeline, thus $u_n^f$ serves as a weighted sum of followings' semantics information according to their relative tweeting frequency.

For followers, as the average quality of followers of an account defines its social status and the quality could be evaluated by its properties, we propose to model the follower relationships as follows:
\begin{equation}
    u_n^t = \frac{1}{|N^t|}\sum_{u\in N^t} r_p(u),
\end{equation}

\noindent where $N^t$ denotes the follower set of a Twitter user, $|\cdot|$ denotes the cardinality of a set and $r_p(u)$ is the property representation of user $u$ generated by the profile-property sub-network.

The following-follower sub-network then produces a raw hidden vector for neighborhood information $u_n = concatenation(u_n^f; u_n^t)$. The intermediate vector is then transformed to produce the Twitter user's neighborhood representation $r_n$:
\begin{equation}
    \label{equ:nend}
    r_n = ReLU(FC_n(u_n)),
\end{equation}

\noindent where $FC_n(\cdot)$ is a fully connected layer and $ReLU(\cdot)$ is the adopted activation function.

\subsection{Co-Influence Aggregator}
\label{subsec:SATARCIA}
So far, we have obtained the representation vectors regarding three and all three aspects of a Twitter user, namely $r_s$, $r_p$ and $r_n$ for tweet semantics, user property and follow relationships. A good bot detector should be comprehensive and robust to tamper. In other words, independently considering each aspect of user information would inevitably jeopardize the robustness of the bot detector. Co-attention has been a successful mechanism at handling correlation between two sequences, but it is not designed for mutual influence between multiple representation vectors. Thus we propose a Co-Influence aggregator to take the mutual correlation between tweet semantics, user property and follow relationships into consideration.

Firstly, the affinity index between a pair of aspects is derived:
\begin{equation}
\label{cobegin}
\begin{aligned}
    F_{sp} = tanh(r_s^T W_{sp} r_p),\\
    F_{pn} = tanh(r_p^T W_{pn} r_n),\\
    F_{ns} = tanh(r_n^T W_{ns} r_s),
\end{aligned}
\end{equation}

\noindent where $W_{sp}$, $W_{pn}$ and $W_{ns}$ are learnable parameters of the aggregator. A hidden representation for each aspect which incorporates relevant information from the other two aspects are derived:
\begin{equation}
\begin{aligned}
    h^s = tanh(W_sr_s + F_{sp}(W_pr_p) + F_{ns}(W_nr_n)),\\
    h^p = tanh(W_pr_p + F_{sp}(W_sr_s) + F_{pn}(W_nr_n)),\\
    h^n = tanh(W_nr_n + F_{ns}(W_sr_s) + F_{pn}(W_pr_p)),
\end{aligned}
\end{equation}

\noindent where $W_s$, $W_p$ and $W_n$ are learnable parameters of the aggregator. Finally, the proposed framework SATAR produces the Twitter user representation $r$ as follows:
\begin{equation}
    \label{coend}
    r =  tanh(W_V\cdot concatenation(h^s;h^p;h^n)),
\end{equation}

\noindent where $W_V$ is a learnable parameter of the aggregator.

\begin{table*}
 \caption{Components of Twitter user information used by each bot detection method.}
 \label{tab:SPN}
 \begin{tabular}{c c c c c c c c c c c} 
 \toprule
  &
  \tabincell{c}{Lee \textit{et}\\ \textit{al.}~\cite{lee2011seven}} &
  \tabincell{c}{Yang\\ \textit{et al.}~\cite{yang2020scalable}} &
  \tabincell{c}{Kudugunta\\ \textit{et al.}~\cite{kudugunta2018deep}} &
  \tabincell{c}{Wei \textit{et}\\ \textit{al.}~\cite{wei2019twitter}} &
  \tabincell{c}{Miller\\ \textit{et al.}~\cite{miller2014twitter}} &
  \tabincell{c}{Cresci\\ \textit{et al.}~\cite{cresci2016dna}} &
  \tabincell{c}{Botometer\\ ~\cite{davis2016botornot}} &
  \tabincell{c}{Alhosseini\\ \textit{et al.}~\cite{ali2019detect}} &
  $\rm SATAR_{FC}$ &
  $\rm SATAR_{FT}$ \\
  \midrule
   
  $\bf Semantic$ &
  \checkmark &  % Lee
  &  % Yang
  \checkmark &  % Kudugunta
  \checkmark &  % Wei
  \checkmark &  % Miller
  \checkmark &  % Cresci
  \checkmark &  % Botometer
  &
  \checkmark &
  \checkmark \\
 
  $\bf Property$ &
  \checkmark &  % Lee
  \checkmark &  % Yang
  \checkmark &  % Kudugunta
  &  % Wei
  \checkmark &  % Miller
  &  % Cresci
  \checkmark &  % Botometer
  \checkmark &
  \checkmark &
  \checkmark \\
   
  $\bf Neighbor$ &
  &  % Lee
  &  % Yang
  &  % Kudugunta
  &  % Wei
  &  % Miller
  &  % Cresci
  \checkmark &  % Botometer
  \checkmark &
  \checkmark &
  \checkmark \\

 \bottomrule
\end{tabular}
\end{table*}
 
\begin{table*}
 \caption{Performance comparison for bot detection methods. “/” denotes insufficient user information to support the baseline.}
 \label{tab:TwiBotMetric}
 \begin{tabular}{c c c c c c c c c c c c c c c c c c c c c c c c c c c c c c c c}
 \toprule
    \multicolumn{2}{c}{} &
    \multicolumn{3}{c}{\tabincell{c}{Lee \textit{et}\\ \textit{al.}~\cite{lee2011seven}}} &
    \multicolumn{3}{c}{\tabincell{c}{Yang\\ \textit{et al.}~\cite{yang2020scalable}}} &
    \multicolumn{3}{c}{\tabincell{c}{Kudugunta\\ \textit{et al.}~\cite{kudugunta2018deep}}} &
    \multicolumn{3}{c}{\tabincell{c}{Wei \textit{et} \\\textit{al.}~\cite{wei2019twitter}}} &
    \multicolumn{3}{c}{\tabincell{c}{Miller\\ \textit{et al.}~\cite{miller2014twitter}}} &
    \multicolumn{3}{c}{\tabincell{c}{Cresci\\ \textit{et al.}~\cite{cresci2016dna}}} &
    \multicolumn{3}{c}{\tabincell{c}{\tabincell{c}{Botometer\\ ~\cite{davis2016botornot}}}} &
    \multicolumn{3}{c}{\tabincell{c}{Alhosseini\\ \textit{et al.}~\cite{ali2019detect}}} &
    \multicolumn{3}{c}{$\rm SATAR_{FC}$} & \multicolumn{3}{c}{$\rm SATAR_{FT}$} \\

    % & & \textbf{S} & \textbf{P} & \textbf{N} & \textbf{S} & \textbf{P} & \textbf{N} & \textbf{S} & \textbf{P} & \textbf{N} & \textbf{S} & \textbf{P} & \textbf{N} &\textbf{S} & \textbf{P} & \textbf{N} &\textbf{S} & \textbf{P} & \textbf{N} &\textbf{S} & \textbf{P} & \textbf{N} &\textbf{S} & \textbf{P} & \textbf{N} &\textbf{S} & \textbf{P} & \textbf{N} &\textbf{S} & \textbf{P} & \textbf{N} \\
     
    %& & \checkmark &   & \checkmark & \checkmark & & & \checkmark & & & \checkmark & \checkmark & & & \checkmark &  &  & \checkmark & & \checkmark & \checkmark & \checkmark & \checkmark & \checkmark & \checkmark & \checkmark & \checkmark & \checkmark & \checkmark & \checkmark & \checkmark \\ 
 \midrule
 
 % \multicolumn{2}{c}{$\bf Semantic$} & \multicolumn{3}{c}{\checkmark} & \multicolumn{3}{c}{\checkmark} & \multicolumn{3}{c}{\checkmark} & \multicolumn{3}{c}{\checkmark} & \multicolumn{3}{c}{\checkmark} & & & & \multicolumn{3}{c}{\checkmark} & & & & \multicolumn{3}{c}{\checkmark} & \multicolumn{3}{c}{\checkmark} \\
 
  % \multicolumn{2}{c}{$\bf Property$} & \multicolumn{3}{c}{\checkmark} & \multicolumn{3}{c}{\checkmark} & & & & & & & \multicolumn{3}{c}{\checkmark} & \multicolumn{3}{c}{\checkmark} & \multicolumn{3}{c}{\checkmark} & \multicolumn{3}{c}{\checkmark} & \multicolumn{3}{c}{\checkmark} & \multicolumn{3}{c}{\checkmark} \\
   
 %  \multicolumn{2}{c}{$\bf Neighbor$} & & & & & & & & & & & & & & & & & & & \multicolumn{3}{c}{\checkmark} & \multicolumn{3}{c}{\checkmark} & \multicolumn{3}{c}{\checkmark} & \multicolumn{3}{c}{\checkmark} \\
 
 % \midrule

 \multirow{3}{*}{\textbf{TwiBot-20}} &
 Acc & 
 \multicolumn{3}{c}{0.7456} &
 \multicolumn{3}{c}{0.8191} &
 \multicolumn{3}{c}{0.8174} &
 \multicolumn{3}{c}{0.7126} &
 \multicolumn{3}{c}{0.4801} &
 \multicolumn{3}{c}{0.4793} &
 \multicolumn{3}{c}{0.5584} & 
 \multicolumn{3}{c}{0.6813} & 
 \multicolumn{3}{c}{0.7838} & 
 \multicolumn{3}{c}{\bf 0.8412} \\
 
%  & Precision & \multicolumn{3}{c}{0.6477} & \multicolumn{3}{c}{0.5124} &  \multicolumn{3}{c}{0.7281} & \multicolumn{3}{c}{0.7033} & \multicolumn{3}{c}{\bf 1.0000} & \multicolumn{3}{c}{0.7560} & \multicolumn{3}{c}{0.6527} & \multicolumn{3}{c}{0.6706} & \multicolumn{3}{c}{0.7770} & \multicolumn{3}{c}{0.8049} \\
 
%   & Recall & \multicolumn{3}{c}{0.8453} & \multicolumn{3}{c}{0.8063} &  \multicolumn{3}{c}{0.0578} & \multicolumn{3}{c}{0.8109} & \multicolumn{3}{c}{0.6022} & \multicolumn{3}{c}{\bf 0.9828} & \multicolumn{3}{c}{0.3912} & \multicolumn{3}{c}{0.8053} & \multicolumn{3}{c}{0.8424} & \multicolumn{3}{c}{0.9329} \\
  
%  & specificity & \multicolumn{3}{c}{0.6340} & \multicolumn{3}{c}{0.9761} & \multicolumn{3}{c}{0.5967} & \multicolumn{3}{c}{\bf 1.0000} & \multicolumn{3}{c}{0.6041} & \multicolumn{3}{c}{0.6262} & \multicolumn{3}{c}{0.7551} & \multicolumn{3}{c}{0.6051} & \multicolumn{3}{c}{0.7145} & \multicolumn{3}{c}{0.7330} \\
 
 &
 F1&
 \multicolumn{3}{c}{0.7823} &
 \multicolumn{3}{c}{0.8546} &
 \multicolumn{3}{c}{0.7517} &
 \multicolumn{3}{c}{0.7533} &
 \multicolumn{3}{c}{0.6266} &
 \multicolumn{3}{c}{0.1072} &
 \multicolumn{3}{c}{0.4892} &
 \multicolumn{3}{c}{0.7318} &
 \multicolumn{3}{c}{0.8084} &
 \multicolumn{3}{c}{\bf 0.8642} \\

 &
 MCC &
 \multicolumn{3}{c}{0.4879} &
 \multicolumn{3}{c}{0.6643} &
 \multicolumn{3}{c}{0.6710} &
 \multicolumn{3}{c}{0.4193} & 
 \multicolumn{3}{c}{-0.1372} & 
 \multicolumn{3}{c}{0.0839} & 
 \multicolumn{3}{c}{0.1558} & 
 \multicolumn{3}{c}{0.3543} & 
 \multicolumn{3}{c}{0.5637} & 
 \multicolumn{3}{c}{\bf 0.6863} \\
 
%  \bottomrule

% \end{tabular}
% \end{table*}

% \begin{table*}
%  \caption{Performance Comparison for Bot Detection Methods on cresci-17}
%  \label{tab:CresciMetric}
%  \begin{tabular}{c c c c c c c c c c c c c c c c c c c c c c c c c c c c c c c c} 
%  \toprule
 
%   & & \multicolumn{3}{c}{$\bf RF$}& \multicolumn{3}{c}{$\bf StreamCluster$} & \multicolumn{3}{c}{$\bf DNA$} & \multicolumn{3}{c}{$\bf LSTM$} & \multicolumn{3}{c}{$\bf LSTM_P$}  & \multicolumn{3}{c}{$\bf Property^{+}$} & \multicolumn{3}{c}{$\bf Botometer$} & \multicolumn{3}{c}{$\bf GCN$} & \multicolumn{3}{c}{$\bf SATAR_{FC}$} & \multicolumn{3}{c}{$\bf SATAR_{FT}$} \\

 \midrule
 \multirow{3}{*}{\textbf{Cresci-17}} & 
 Acc & 
 \multicolumn{3}{c}{0.9750} & 
 \multicolumn{3}{c}{0.9847} & 
 \multicolumn{3}{c}{0.9799} & 
 \multicolumn{3}{c}{0.9670} & 
 \multicolumn{3}{c}{0.5204} & 
 \multicolumn{3}{c}{0.4029} & 
 \multicolumn{3}{c}{0.9597} & 
 \multicolumn{3}{c}{/} & 
 \multicolumn{3}{c}{0.9622} & 
 \multicolumn{3}{c}{\bf 0.9871} \\
 
%  & Precision & \multicolumn{3}{c}{0.9864}& \multicolumn{3}{c}{0.3490} & \multicolumn{3}{c}{0.9871} & \multicolumn{3}{c}{0.9863} & \multicolumn{3}{c}{0.9655}  & \multicolumn{3}{c}{0.9932} & \multicolumn{3}{c}{0.9731} & \multicolumn{3}{c}{/} & \multicolumn{3}{c}{0.9764} & \multicolumn{3}{c}{\bf 0.9955} \\
 
%   & Recall & \multicolumn{3}{c}{0.9787} & \multicolumn{3}{c}{0.7371}& \multicolumn{3}{c}{0.1715} & \multicolumn{3}{c}{0.9675} & \multicolumn{3}{c}{0.9628}  & \multicolumn{3}{c}{0.9854} & \multicolumn{3}{c}{0.9731} & \multicolumn{3}{c}{/} & \multicolumn{3}{c}{0.9709} & \multicolumn{3}{c}{\bf 0.9866} \\
  
%  & specificity & \multicolumn{3}{c}{/} & \multicolumn{3}{c}{\bf 0.9943} & \multicolumn{3}{c}{0.9656} & \multicolumn{3}{c}{0.9865} & \multicolumn{3}{c}{0.9885} & \multicolumn{3}{c}{0.9828} & \multicolumn{3}{c}{0.9195} & \multicolumn{3}{c}{0.9114} & \multicolumn{3}{c}{0.9398} & \multicolumn{3}{c}{0.9885} \\
 
 &
 F1&
 \multicolumn{3}{c}{0.9826} &
 \multicolumn{3}{c}{0.9893} &
 \multicolumn{3}{c}{0.9641} & 
 \multicolumn{3}{c}{0.9768} & 
 \multicolumn{3}{c}{0.4737} & 
 \multicolumn{3}{c}{0.2923} & 
 \multicolumn{3}{c}{0.9731} & 
 \multicolumn{3}{c}{/} & 
 \multicolumn{3}{c}{0.9737} & 
 \multicolumn{3}{c}{\bf 0.9910} \\

 &
 MCC &
 \multicolumn{3}{c}{0.9387} &
 \multicolumn{3}{c}{0.9625} &
 \multicolumn{3}{c}{0.9501} &
 \multicolumn{3}{c}{0.9200} &
 \multicolumn{3}{c}{0.1573} &
 \multicolumn{3}{c}{0.2255} &
 \multicolumn{3}{c}{0.8926} &
 \multicolumn{3}{c}{/} &
 \multicolumn{3}{c}{0.9069} &
 \multicolumn{3}{c}{\bf 0.9685} \\

%   \bottomrule
% \end{tabular}
% \end{table*}

% \begin{table*}
%  \caption{Performance Comparison for Bot Detection Methods on PAN-19}
%  \label{tab:PANMetric}
%  \begin{tabular}{c c c c c c c c c c c c c c c c c c c c c c c c c c c c c c c c} 
%  \toprule
 
%   & & \multicolumn{3}{c}{$\bf RF$}& \multicolumn{3}{c}{$\bf StreamCluster$} & \multicolumn{3}{c}{$\bf DNA$} & \multicolumn{3}{c}{$\bf LSTM$} & \multicolumn{3}{c}{$\bf LSTM_P$}  & \multicolumn{3}{c}{$\bf Property^{+}$} & \multicolumn{3}{c}{$\bf Botometer$} & \multicolumn{3}{c}{$\bf GCN$} & \multicolumn{3}{c}{$\bf SATAR_{FC}$} & \multicolumn{3}{c}{$\bf SATAR_{FT}$} \\
 
 \midrule 
 \multirow{3}{*}{\textbf{PAN-19}} &
 Acc &
 \multicolumn{3}{c}{/} &
 \multicolumn{3}{c}{/} &
 \multicolumn{3}{c}{/} &
 \multicolumn{3}{c}{0.9464} &
 \multicolumn{3}{c}{/} &
 \multicolumn{3}{c}{0.8797} &
 \multicolumn{3}{c}{/} &
 \multicolumn{3}{c}{/} &
 \multicolumn{3}{c}{0.8728} &
 \multicolumn{3}{c}{\bf 0.9509} \\
 
%  & Precision & \multicolumn{3}{c}{/}& \multicolumn{3}{c}{/} & \multicolumn{3}{c}{0.9483} & \multicolumn{3}{c}{ \bf 0.9775} & \multicolumn{3}{c}{/}  & \multicolumn{3}{c}{/} & \multicolumn{3}{c}{/} & \multicolumn{3}{c}{/} & \multicolumn{3}{c}{0.8737} & \multicolumn{3}{c}{0.9494} \\
 
%   & Recall & \multicolumn{3}{c}{/}& \multicolumn{3}{c}{/} & \multicolumn{3}{c}{0.8039} & \multicolumn{3}{c}{0.9142} & \multicolumn{3}{c}{/}  & \multicolumn{3}{c}{/} & \multicolumn{3}{c}{/} & \multicolumn{3}{c}{/} & \multicolumn{3}{c}{0.8722} & \multicolumn{3}{c}{\bf 0.9527} \\
  
%  & specificity & \multicolumn{3}{c}{/} & \multicolumn{3}{c}{0.9560} & \multicolumn{3}{c}{ \bf 0.9789} & \multicolumn{3}{c}{/} & \multicolumn{3}{c}{/} & \multicolumn{3}{c}{/} & \multicolumn{3}{c}{/} & \multicolumn{3}{c}{0.8848} & \multicolumn{3}{c}{0.8734} & \multicolumn{3}{c}{0.9490} \\
 
 &
 F1&
 \multicolumn{3}{c}{/} &
 \multicolumn{3}{c}{/} &
 \multicolumn{3}{c}{/} &
 \multicolumn{3}{c}{0.9448} &
 \multicolumn{3}{c}{/} &
 \multicolumn{3}{c}{0.8701} &
 \multicolumn{3}{c}{/} &
 \multicolumn{3}{c}{/} &
 \multicolumn{3}{c}{0.8729} &
 \multicolumn{3}{c}{\bf 0.9510} \\

 &
 MCC &
 \multicolumn{3}{c}{/} &
 \multicolumn{3}{c}{/} &
 \multicolumn{3}{c}{/} &
 \multicolumn{3}{c}{0.8948} &
 \multicolumn{3}{c}{/} &
 \multicolumn{3}{c}{0.7685} &
 \multicolumn{3}{c}{/} &
 \multicolumn{3}{c}{/} &
 \multicolumn{3}{c}{0.7456} &
 \multicolumn{3}{c}{\bf 0.9018} \\

 \bottomrule
\end{tabular}
\end{table*}

\begin{algorithm}[!t]
    \caption{SATAR Learning Algorithm}
    \label{alg:SATAR}
    \SetAlgoLined
    \KwIn{Twitter user dataset $TU$, each user $u \in TU$ has tweets $T$, properties $P$ and neighbors $N$}
    \KwOut{SATAR-optimized parameters $\theta$}
    Initialize $\theta$; \\
    \For{each user $u \in TU$}
    {
    Initialize $r_n(u)$; \\
    $u.y \leftarrow$ self-supervised label assignment according to user $u$'s follower count; \\
    }
    \While{$\theta$ has not converged}
    {
    \For{each user $u \in TU$}
    {
    $r_s(u) \leftarrow$ Equation (\ref{sbegin} - \ref{send}) with $u.T$; \\
    $r_p(u) \leftarrow$ Equation (\ref{p}) with $u.P$; \\
    $r(u) \leftarrow$ Equation (\ref{cobegin} - \ref{coend}) with $r_s(u)$, $r_p(u)$ and $r_n(u)$; \\
    $L_u \leftarrow$ Equation (\ref{lossbegin} - \ref{equ:lossend}) with $r(u)$ and $u.y$; \\
    }
    $\theta \leftarrow$ BackPropagate($L_u$); \\
    \For{each user $u \in TU$}
    {
    $r_n(u) \leftarrow$ Equation (\ref{nbegin} - \ref{equ:nend}) with $u.N$;
    }
    }
\end{algorithm}

\subsection{Self-Supervised Learning and Optimization}
\label{subsec:SATARSSLO}
Twitter user representation learning attempts to model a specific user with a distributed representation. We adopt \textbf{follower count} as the self-supervised signal for SATAR training. Specifically, a user's follower count is separated into several categories based on its numerical scale and the overall follower count distribution. We train the representation learning framework SATAR to classify each user into such categories, obtaining user representation in the process. We believe that \textbf{follower count} would be an ideal self-supervised training signal due to the following reasons:

\begin{itemize}[leftmargin=*]
    \item Self-supervised training with follower count is task-agnostic. Whether it is bot detection, content recommendation or online campaign modeling, follower count relates to all tasks on social media without being specific to any of them.
    \item Follower count is most representative of a Twitter user. There is no better choice to describe a Twitter user more efficiently and accurately, especially when follower count also involves the evaluation of other users.
    \item Follower count is more robust to large-scale tamper. Although it is possible to purchase fake followers, according to Cresci \textit{et al.}~\cite{Cresci2015FameFS}'s investigation, an increase of 1,000 followers often costs from 13 to 19 U.S. dollars. As a result, it is costly to significantly alter the magnitude of a user's follower count, let alone launch a campaign with many active bots.
\end{itemize}

Specifically, assuming that a user could be categorized into $D$ classes based on its follower count, a softmax layer is applied to the representation of the user $r$:

\begin{equation}
    \label{lossbegin}
    \hat{y} = softmax(W_fr + b_f),
\end{equation}

\noindent where $\hat{y} = [\hat{y_1}, \hat{y_2}, \cdot \cdot \cdot, \hat{y_D}]$ is the predicted probability vector for each class, $W_f$ and $b_f$ are learnable parameters. $y = [y_1,y_2,\cdot \cdot \cdot, y_D]$ denotes the self-supervised ground-truth for such classification in one-hot encoding. We minimize the cross-entropy loss function as follows:
\begin{equation}
    \label{equ:lossend}
    L(\theta) = -\sum_{1 \leqslant i \leqslant D} y_i log(\hat{y_i}),
\end{equation}

\noindent where $\theta$ denotes the parameters in the proposed framework SATAR.

% For SATAR's application in Twitter bot detection, we firstly pre-train the framework with a vast amount of self-supervised users, then replaces the rear softmax layer with
% \begin{equation}
%     y_b = softmax(vW_b + b_b)
% \end{equation}

% \noindent where $y_b = [y_b^0, y_b^1]$ is the predicted probability vector with $y_b^0$ as the probability of a genuine user and $y_b^1$ as the probability of a Twitter bot. $y \in \{0,1\}$ denotes the ground truth label of users. For each user, the goal is to minimize the cross-entropy loss:
% \begin{equation}
%     L(\theta) = -ylog(y_v^1) - (1-y)log(1-y_v^1)
% \end{equation}

% \noindent where $\theta$ denotes the parameters in the proposed framework SATAR.

Algorithm \ref{alg:SATAR} presents the overall training schema of our proposed Twitter account representation learning framework SATAR.

\section{Experiments}  % read 4 kdd-2019 articles and the titles are "Experiments"
\label{sec:experiments}
In this section, we conduct extensive experiments with in-depth analysis on three real-world bot detection datasets.

\subsection{Experiment Settings}
\label{subsec:expsetting}
In this section, we provide information about datasets, bot detection baselines and evaluation metrics adopted in the experiments. 

\noindent \textbf{Datasets.} We make use of three datasets, {\verb|TwiBot-20|}, {\verb|cresci-17|} and {\verb|PAN-19|}. 
% They are of varying information completeness and collection time. 
As Twitter bots bear different purposes and evolve rapidly, these high quality datasets are adopted to provide a comprehensive evaluation and verify the generalizability and adaptability of baselines and our proposed method.

\begin{itemize} [leftmargin=*]

\item {\verb|TwiBot-20|}~\cite{feng2021twibot} is a comprehensive sample of the current Twittersphere to evaluate whether bot detection methods can generalize in real-world scenarios. Users in {\verb|TwiBot-20|} could be generally split into four interest domains: politics, business, entertainment and sports. As of user information, {\verb|TwiBot-20|} contains semantic, property and neighborhood information of Twitter users.

% We create a new dataset {\verb|TwiBot-20|}. Section 3 explains the methodology for data collection and annotation of dataset ours.

\item {\verb|cresci-17|} ~\cite{cresci2017paradigm} is a public dataset with 4 components: genuine accounts, social spambots, traditional spambots and fake followers. We merge the four parts and utilize {\verb|cresci-17|} as a whole. {\verb|cresci-17|} contains semantic and property information.
% , such as mayoral election in Rome in 2014
% , Twitter accounts which use the {\verb|#TALNTS|} hashtag, and accounts which advertised products on sale on Amazon\footnote{https://www.amazon.com/}

\item {\verb|PAN-19|} \footnote{\url{https://zenodo.org/record/3692340}} is a dataset of a Bots and Gender Profiling shared task in the PAN workshop at CLEF 2019. It is used for bots and gender profiling and only contains user semantic information.
\end{itemize}

% \begin{table}
% \setlength{\tabcolsep}{1pt}
% \setlength{\abovecaptionskip}{1pt}
% \caption{Overview of three adopted bot detection datasets.}
% \label{tab:dataset}
% \begin{tabular}{c c c c c c c c}
% \toprule

% Dataset & \tabincell{c}{User \\ Count} & \tabincell{c}{Human \\ Count} & \tabincell{c}{Bot \\ Count} & \tabincell{c}{S \\ Info} & \tabincell{c}{P \\ Info} & \tabincell{c}{N \\ Info} & \tabincell{c}{Release\\ Year} \\ \midrule

% \ \ TwiBot-20\ \ & \ \ 229,573 \ \ & \ \ 5,237 \ \ & \ \ 6,589 \ \ &  \ \ \checkmark \ \  & \ \  \checkmark \ \  & \ \ \checkmark \ \  & \ \ 2020 \ \  \\

% Cresci-17 & 9,813 & 2,764 & 7,049 & \checkmark & \checkmark & & 2017 \\

% PAN-19 & 11,378 & 5,765 & 5,613 & \checkmark & & & 2019 \\
 
% \bottomrule
% \end{tabular}

% \end{table}

\begin{table}
\setlength{\tabcolsep}{1pt}
\setlength{\abovecaptionskip}{1pt}
\caption{Overview of three adopted bot detection datasets.}
\label{tab:dataset}
\begin{tabular}{c c c c}
\toprule

Dataset & \ \ User Count \ \ & \ \ Human Count \ \ & \ \ Bot Count  \ \ \\ \midrule

\ \ TwiBot-20\ \ & \ \ 229,573 \ \ & \ \ 5,237 \ \ & \ \ 6,589 \ \   \\

Cresci-17 & 9,813 & 2,764 & 7,049  \\

PAN-19 & 11,378 & 5,765 & 5,613  \\
 
\bottomrule
\end{tabular}

\end{table}

A summary of these three datasets is presented in Table~\ref{tab:dataset}. We randomly conduct a 7:2:1 partition for three datasets as training, validation and test set. Such a partition is shared across all experiments in Section ~\ref{subsec:expBDP}, Section ~\ref{subsec:GeneralizeStudy} and Section ~\ref{subsec:AdaptStudy}. We choose these three benchmarks out of numerous bot detection datasets due to their larger size, collection time span and superior annotation quality.

\noindent \textbf{Baseline Methods.} We compare SATAR with the following bot detection methods as baselines:

\begin{figure}[!t]
     \centering 
     \setlength{\belowcaptionskip}{-0.4cm}
     \subfigure[train on politics domain]{\label{fig:subfig:a}
     \includegraphics[width=0.47\linewidth]{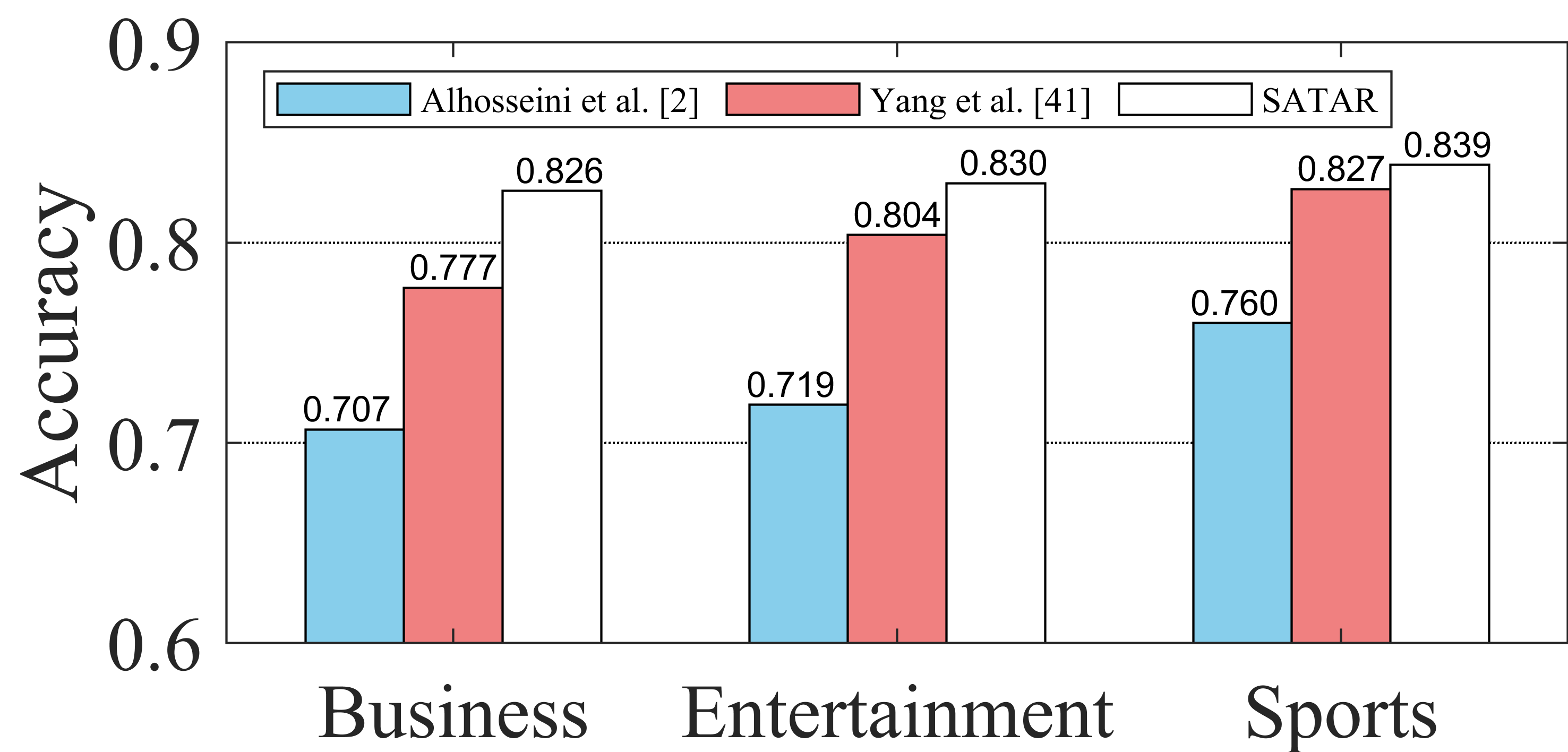}}
     \hspace{0.01\linewidth}
     \subfigure[train on business domain]{\label{fig:subfig:b}
     \includegraphics[width=0.47\linewidth]{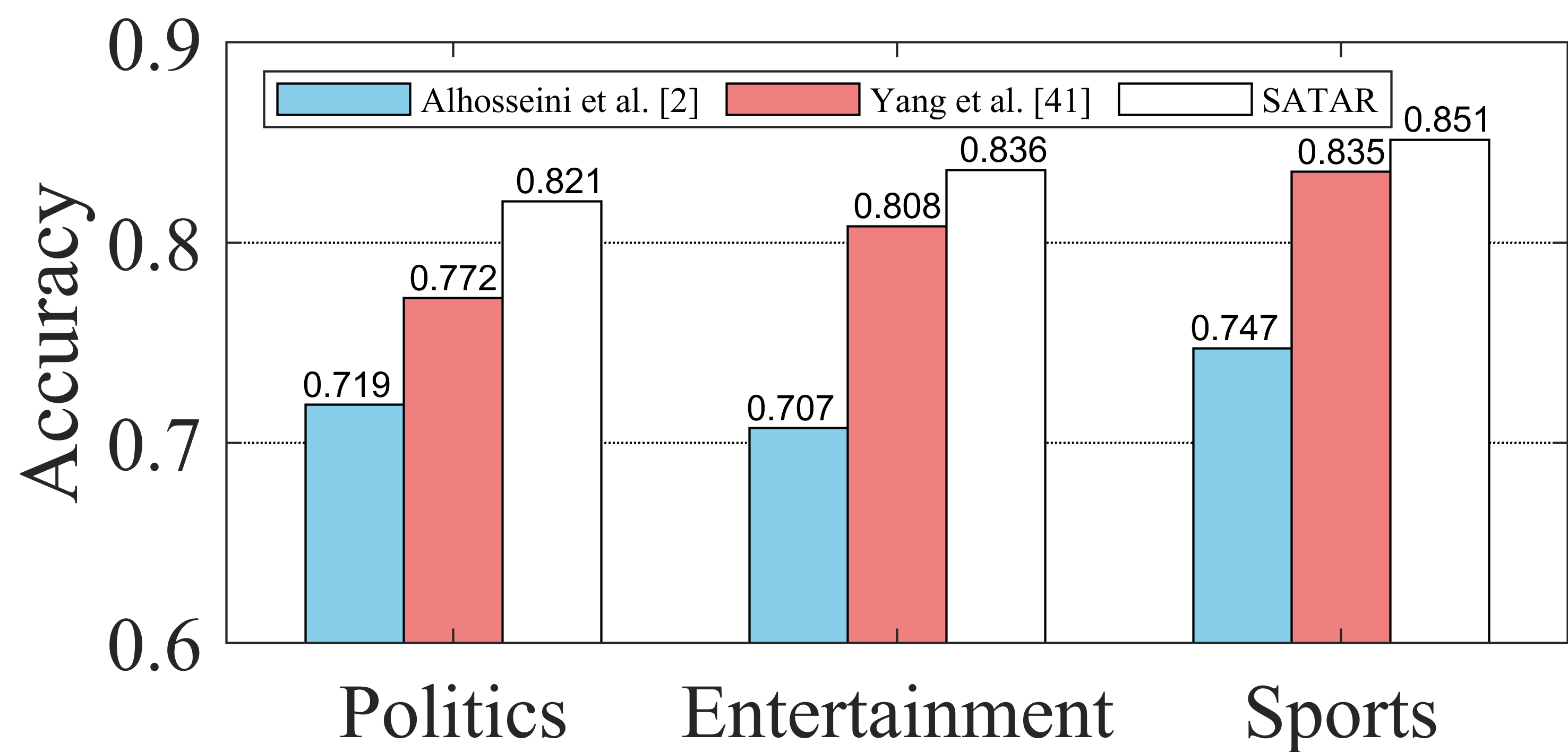}}
     \vfill
     \subfigure[train on entertainment domain]{\label{fig:subfig:c}
     \includegraphics[width=0.47\linewidth]{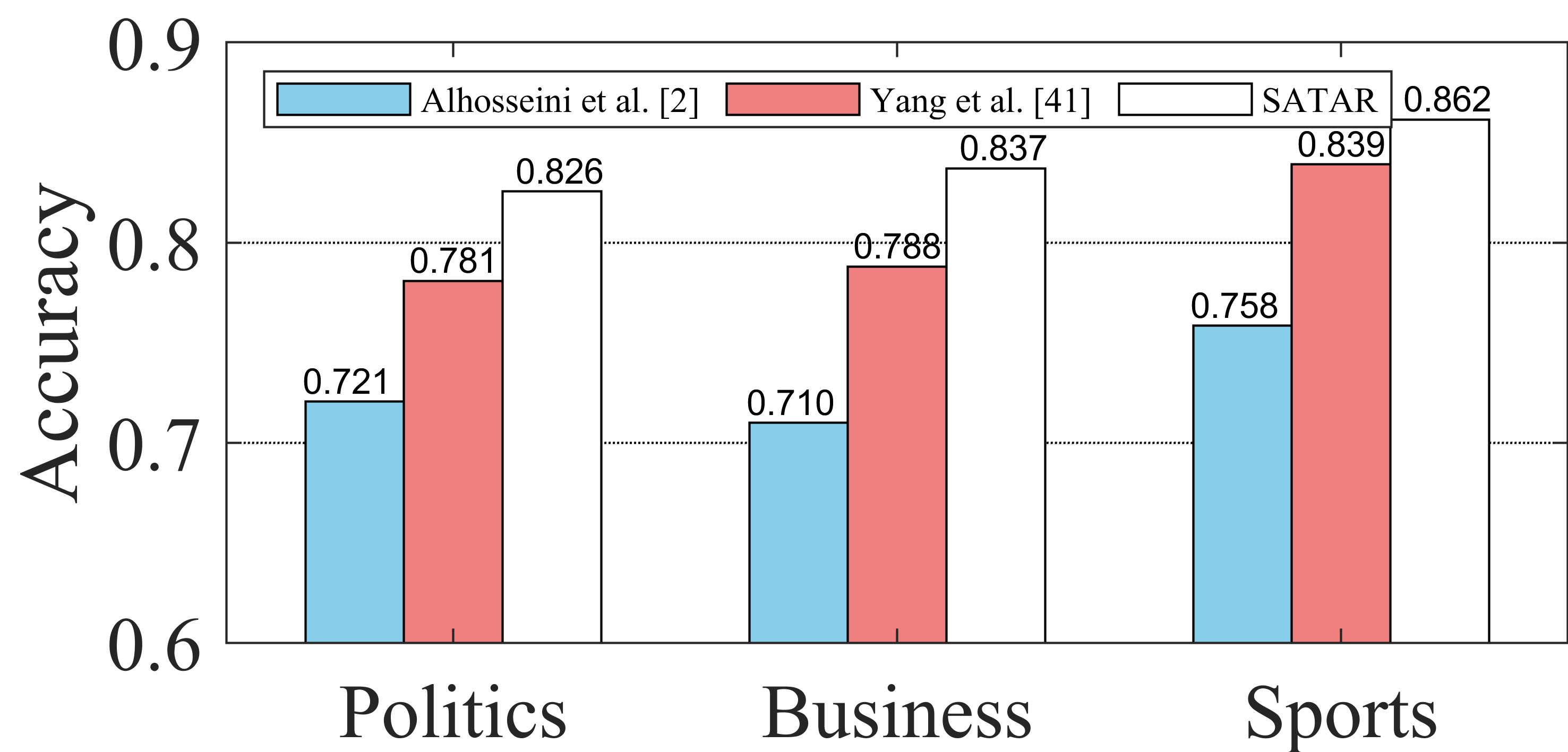}}
     \hspace{0.01\linewidth}
     \subfigure[train on sports domain]{\label{fig:subfig:d}
     \includegraphics[width=0.47\linewidth]{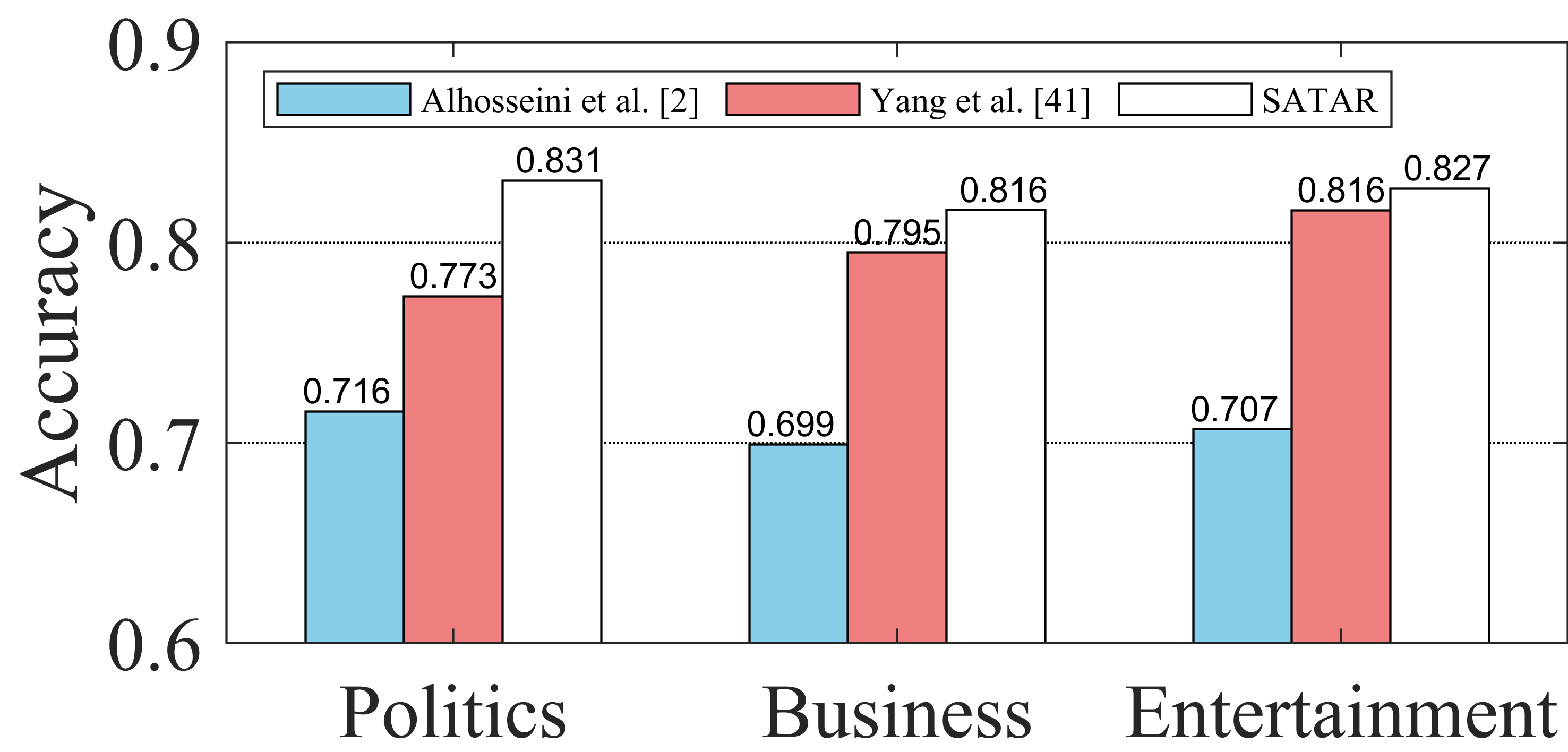}}
     \caption{Train SATAR and two competitive baselines on one domain of TwiBot-20 and test on the other three domains.}
     \label{fig:Domain}
\end{figure}

\begin{itemize} [leftmargin=*]

\item Lee \textit{et al.}~\cite{lee2011seven}: Lee \textit{et al.} use random forest classifier with several Twitter user features. e.g. the longevity of the account.

\item Yang \textit{et al.}~\cite{yang2020scalable}: Yang \textit{et al.} use random forest with minimal account metadata and 12 derived features.

\item Kudugunta \textit{et al.}~\cite{kudugunta2018deep}: Kudugunta \textit{et al.} propose an architecture that uses both tweet content and the metadata. 

\item Wei \textit{et al.}~\cite{wei2019twitter}: Wei \textit{et al.} use word embeddings and a three-layer BiLSTM to encode tweets. A fully connected softmax layer is adopted for binary classification.

\item Miller \textit{et al.}~\cite{miller2014twitter}: Miller \textit{et al.} extract 107 features from a user's tweet and property information. Bot users are conceived as abnormal outliers and modified stream clustering algorithm is adopted to identify Twitter bots.

\item Cresci \textit{et al.}~\cite{cresci2016dna}: Cresci \textit{et al.} utilize strings to represent the sequence of a user's online actions. Each action type can be encoded with a character. By identifying the group of accounts that share the longest common substring, a set of bot accounts are obtained.

\item Botometer~\cite{davis2016botornot}: Botometer 
is a publicly available service that leverages more than one thousand features to classify an account.

\item Alhosseini \textit{et al.}~\cite{ali2019detect}: Alhosseini \textit{et al.} utilize graph convolutional network to detect Twitter bots. It uses following information and user features to learn representations and classify Twitter users.
\end{itemize}

For the following SATAR-based bot detection methods, the self-supervised representation learning step adopts the Pareto Principle\footnote{\url{https://en.wikipedia.org/wiki/Pareto\_principle}} as a self-supervised classification task, where the framework learns to predict whether a Twitter user's follower count is among the top $20\%$ or the bottom $80\%$. It is an instance of the self-supervised representation learning strategy in Section ~\ref{subsec:SATARSSLO}.

\begin{itemize}[leftmargin=*]
\item $\rm SATAR_{FC}$: The proposed representation learning framework SATAR is firstly trained with self-supervised user classification tasks based on their follower count, then the final softmax layer is reinitialized and trained on the task of bot detection.

\item $\rm SATAR_{FT}$: The proposed representation learning framework SATAR is firstly trained using self-supervised users, then the final softmax layer is reinitialized and fine-tuning is performed on the whole framework using the training set of bot detection.
\end{itemize}

\noindent \textbf{Evaluation Metrics.}
We adopt Accuracy, F1-score and MCC~\cite{matthews1975comparison} as evaluation metrics of different bot detection methods. Accuracy is a straightforward indicator of classifier correctness, while F1-score and MCC are more balanced evaluation metrics.

\begin{figure}
    \centering
    \includegraphics[width = \linewidth]{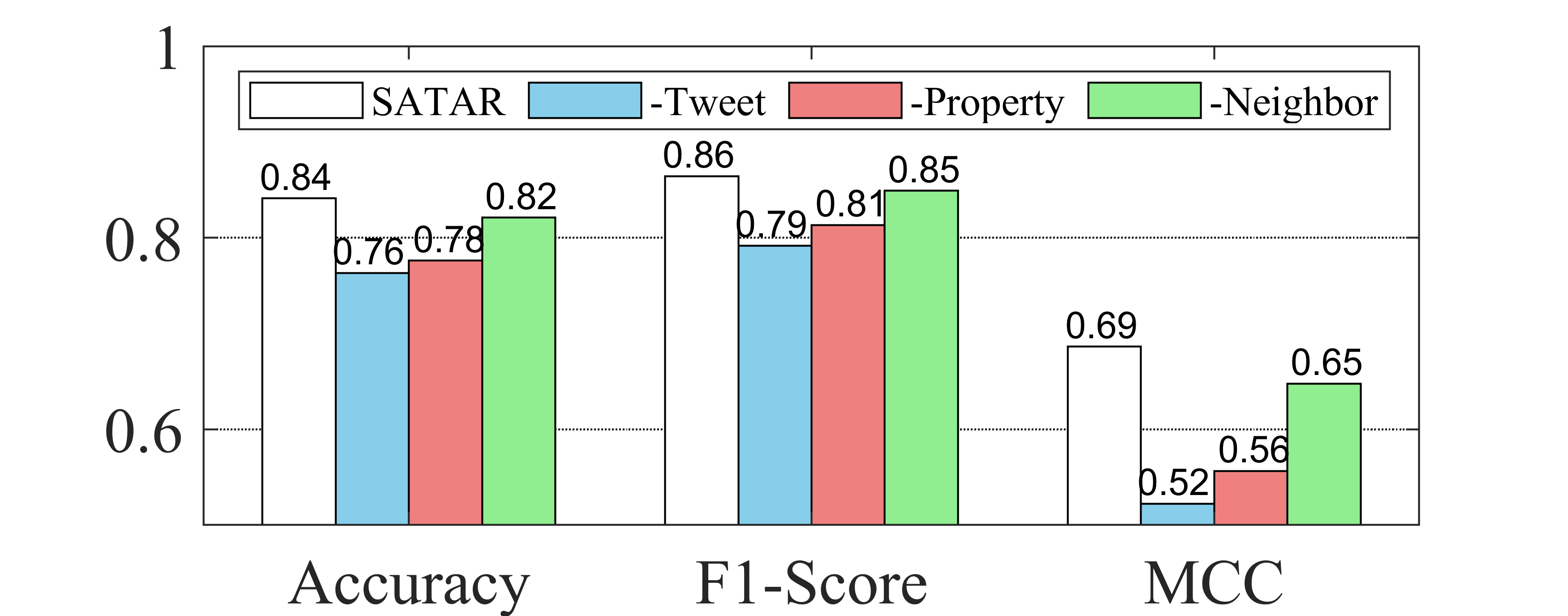}
    \caption{Ablation study that removes the semantic, property and neighborhood sub-networks from SATAR.}
    \label{fig:ablation}
\end{figure}

 \begin{figure*}
    \centering
    \setlength{\belowcaptionskip}{-0.3cm}
    \includegraphics[width = .95\linewidth]{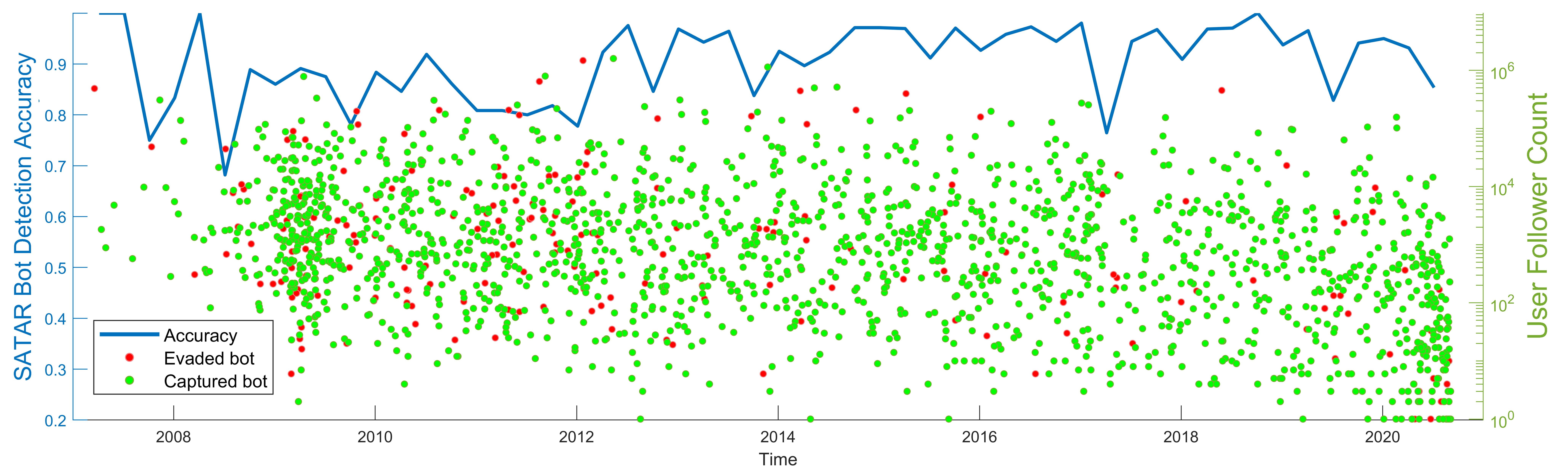}
    \caption{SATAR's prediction of specific users in TwiBot-20. Scattered points demonstrate SATAR's prediction for specific users and the line indicates SATAR's overall accuracy of capturing bots registered in a 3-month time span.}
    \label{fig:time}
\end{figure*}

\begin{figure}
    \centering
    \setlength{\belowcaptionskip}{-0.5cm}
    \includegraphics[width = 0.95\linewidth]{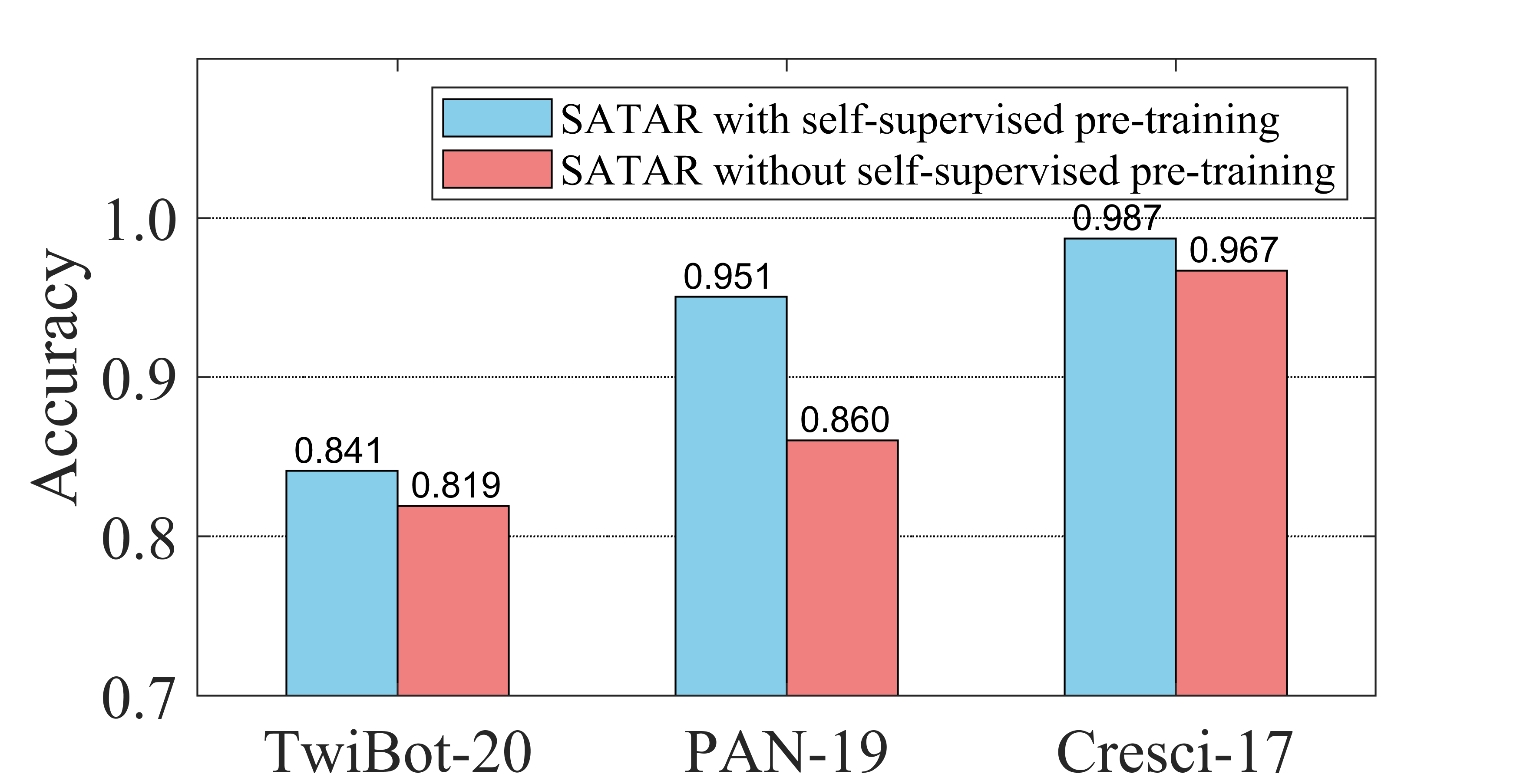}
    \caption{Ablation study removing the self-supervised pre-training step from SATAR and train on the three datasets.}
    \label{fig:super}
    %\vspace{-10pt}
\end{figure}

\subsection{Bot Detection Performance}
\label{subsec:expBDP}
Table \ref{tab:SPN} identifies the user information that each compared method uses. Table ~\ref{tab:TwiBotMetric} reports bot detection performance of different methods on three datasets. Table ~\ref{tab:TwiBotMetric} demonstrates that:

\begin{itemize} [leftmargin=*]
 
 \item $\rm SATAR$ based methods achieve competitive performance compared with other baselines, which demonstrates that SATAR is generally effective in Twitter bot detection. $\rm SATAR_{FT}$ outperforms $\rm SATAR_{FC}$, which demonstrates the efficacy of the pre-training and fine-tuning approach.
 
 \item $\rm SATAR_{FT}$ generalizes to real-world scenarios because it outperforms the state-of-the-art methods on the comprehensive and representative dataset {\verb|TwiBot-20|}, which imitates the real-world Twittersphere. Meanwhile, $\rm SATAR_{FT}$ adapts to evolving generations of bots because it achieves the best performance on all three datasets with varying collection time from 2017 to 2020. Section ~\ref{subsec:GeneralizeStudy} and Section ~\ref{subsec:AdaptStudy} will provide further analysis to demonstrate that SATAR successfully addresses the challenges of generalization and adaptation, while critical components and design choices of SATAR are the reasons behind its success. 

 \item For methods mainly based on LSTM, we see that Kudugunta \textit{et al.} ~\cite{kudugunta2018deep} outperforms Wei \textit{et al.} ~\cite{wei2019twitter}. It indicates that Kudugunta \textit{et al.} ~\cite{kudugunta2018deep} can better capture bots by incorporating property items. $\rm SATAR_{FT}$ leverages even more user information than Kudugunta \textit{et al.} ~\cite{kudugunta2018deep} and achieves better performance, which suggests that bot detection methods should incorporate more aspects of user information.
 
\item Feature-engineering based methods, such as Yang \textit{et al.} ~\cite{yang2020scalable}, perform well on {\verb|cresci-17|} but inferior to $\rm SATAR_{FT}$ on {\verb|TwiBot-20|}. This shows that traditional bot detection methods that emphasize feature engineering fail to adapt to new generations of bots.

 \item Both Alhosseini \textit{et al.} ~\cite{ali2019detect} and $\rm SATAR$ use neighborhood information. $\rm SATAR$ based methods outperform Alhosseini \textit{et al.} ~\cite{ali2019detect}, which shows that $\rm SATAR$ better utilizes user neighbors that put Twitter users into their social context.

\end{itemize}

\subsection{SATAR Generalization Study}
\label{subsec:GeneralizeStudy}

The challenge of generalization in social media bot detection demands bot detectors to simultaneously identify bots that attack in many different ways and exploit diversified user information. To prove that SATAR generalizes, we examine SATAR and competitive baselines' performance on {\verb|TwiBot-20|}. As demonstrated in Table ~\ref{tab:TwiBotMetric}, SATAR outperforms all baselines on {\verb|TwiBot-20|}. Given the fact that {\verb|TwiBot-20|} contains diversified bots and human which imitates the real-world Twittersphere, SATAR is demonstrated to best generalize in real-world scenarios. 

To further prove SATAR's generalizability, we train SATAR and two competitive baselines, Alhosseini \textit{et al.} ~\cite{ali2019detect} and Yang \textit{et al.} ~\cite{yang2020scalable}, on one of the four user domains and test on the others. The results are presented in Figure ~\ref{fig:Domain}. It is illustrated that SATAR could better capture other types of bots even when not explicitly trained on them, which further establishes the claim that SATAR successfully generalizes to diversified bots that co-exist on social media.

SATAR is designed to generalize by jointly leveraging all three aspects of user information, namely semantic, property and neighborhood information. To figure out whether our proposal of using as much user information as possible has lead to the generalizability of SATAR, we conduct ablation study that removes one aspect of user information at a time. The results are demonstrated in Figure ~\ref{fig:ablation}.

Results in Figure ~\ref{fig:ablation} show that removing any aspect of information from SATAR would result in a considerable loss in performance, limiting SATAR's ability to generalize to different types of bots. It indicates that SATAR's strategy of leveraging more aspects of information is crucial in its generalization.

\subsection{SATAR Adaptation Study}
\label{subsec:AdaptStudy}

 \begin{figure*}[h]
     \centering
     \includegraphics[width = 0.95\textwidth]{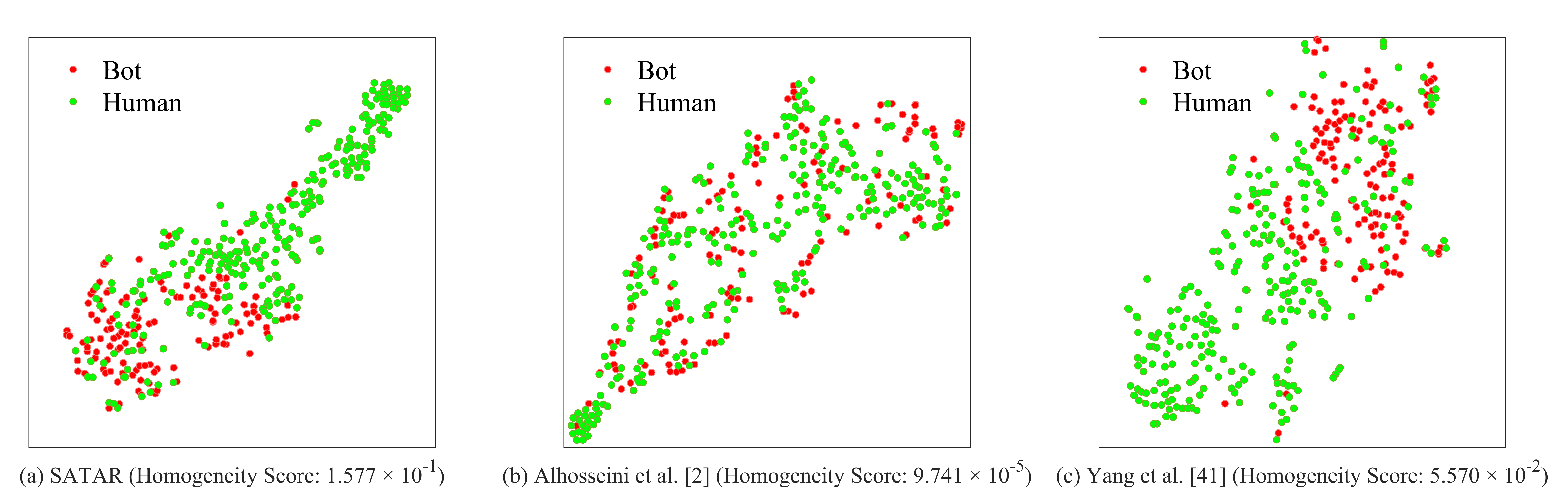}
     \caption{2D t-SNE plot of the user representation vectors of SATAR, Alhosseini \textit{et al.} ~\cite{ali2019detect} and Yang \textit{et al.} ~\cite{yang2020scalable}.}
     \label{fig:representation}
 \end{figure*}

% \vspace{-2pt}

The challenge of adaptation in bot detection demands bot detectors to maintain desirable performance in different times and catch up with rapid bot evolution. To prove that SATAR adapts, we examine SATAR and competitive baselines' performance on three datasets, since they are released in 2017, 2019 and 2020 respectively and could well characterize the bot evolution. Results in Table ~\ref{tab:TwiBotMetric} demonstrate that SATAR reaches state-of-the-art performance on all three datasets, which indicates that SATAR is more successful at adapting to the bot evolution than existing baselines.

To further prove SATAR's ability to adapt, we examine SATAR's prediction of users in dataset {\verb|TwiBot-20|}'s validation set and test set. We present SATAR's prediction results of specific users and SATAR's accuracy in any 3-month time span of user registration time in Figure ~\ref{fig:time}. It is illustrated that SATAR maintains a steady detection accuracy for users created from 2007 to 2020, which further establishes the claim that SATAR successfully adapts to the everlasting bot evolution.

SATAR is designed to adapt by pre-training on mass self-supervised users and fine-tuning on specific bot detection scenarios. To figure out whether this pre-training and fine-tuning schema has enabled SATAR to adapt to newly evolved bots, we conduct ablation study to remove the self-supervised pre-training step. SATAR's performance on different datasets are illustrated in Figure ~\ref{fig:super}.

Figure ~\ref{fig:super} shows that SATAR's performance increases with the adoption of the self-supervised pre-training step, and such trend is especially salient on the dataset PAN-19 with less user information. It indicates that SATAR's ability to adapt indeed comes from the innovative strategy to use follower count as a self-supervised signal for user representation pre-training.

%\vspace{-5pt}
\subsection{Representation Learning Study}
\label{subsec:expRLS}
SATAR improves representation learning for Twitter users. Extrinsic evaluation has proven that SATAR representations are of desirable quality. We further conduct intrinsic evaluation by comparing SATAR representations with Alhosseini \textit{et al.} ~\cite{ali2019detect} and Yang \textit{et al.} ~\cite{yang2020scalable}, which also provide user representations. We cluster representations using $k$-means with $k = 2$, and calculate the homogeneity score, which is the extent to which clusters contain a single class. Higher homogeneity score indicates that users with the same label are more likely to be close to each other.

Figure ~\ref{fig:representation} visualizes representations of users in a subgraph of {\verb|TwiBot-20|}. Figure ~\ref{fig:representation}(a) is the t-SNE plot of SATAR representations, which shows moderate collocation for groups of bot and human, while Figure ~\ref{fig:representation}(b) and (c) show little collocation. Quantitatively, SATAR achieves the highest homogeneity score, which indicates that SATAR produces user representations of higher quality.

% \begin{figure*}[h]
%     \centering
%     \includegraphics[width = \textwidth]{representation.png}
%     \caption{2D t-SNE plot of the user representations of our proposed framework SATAR, Alhosseini \textit{et al.} ~\cite{ali2019detect} and Yang \textit{et al.} ~\cite{yang2020scalable}.}
%     \label{fig:representation}
% \end{figure*}

\begin{figure}
  \centering
  \includegraphics[width=0.4\textwidth]{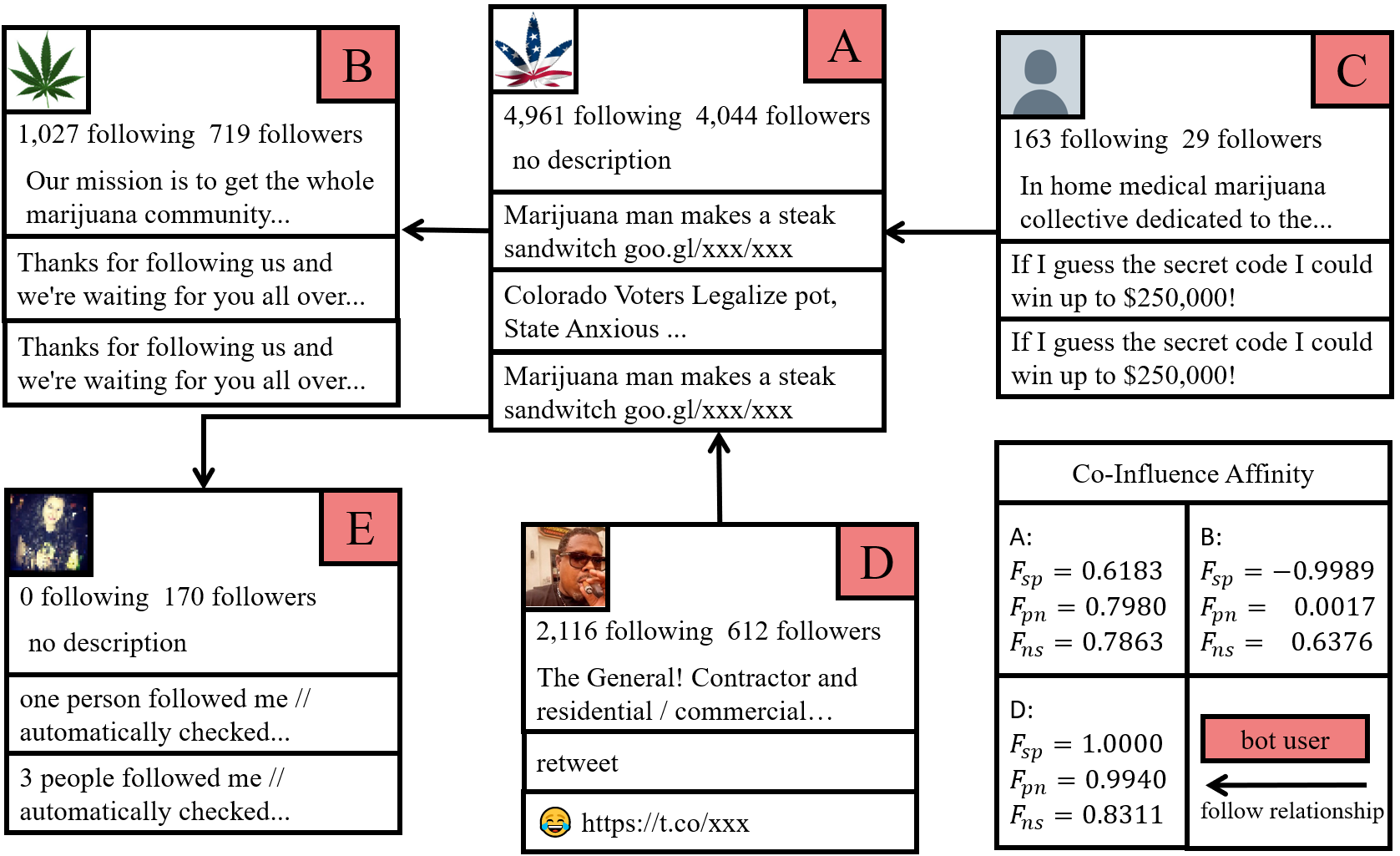}
  \setlength{\belowcaptionskip}{-0.3cm}
  \caption{A sample bot cluster to explain SATAR's decision.}
  \label{fig:case}
\end{figure}

\subsection{Case Study}
To further understand how SATAR identifies bots, we study a specific case of several bots. We use the affinity index values in Equation (\ref{cobegin}) to quantitatively analyze SATAR's decision making. Figure ~\ref{fig:case} shows the detailed information of the sampled users:

\begin{itemize} [leftmargin=*]
    \item SATAR identifies user B and E through their repeated or similar tweets that signal automation. For example, user B has affinity values of $F_{sp} = -0.9989$, $F_{pn} = 0.0017$ and $F_{ns} = 0.6376$. Absolute values of $F_{sp}$ and $F_{ns}$ are significantly greater than $F_{pn}$, which demonstrates that semantic information is the dominant factor for SATAR's decision in this case.
    \item SATAR identifies user C and D through their properties. Abnormal characteristics such as too many followings and default background image are detected by SATAR. User D has larger absolute values for $F_{sp}$ and $F_{pn}$ than $F_{ns}$, which shows that property information is critical in SATAR's judgement.
    \item SATAR captures that user A has four bots as neighbors, which is unlikely for genuine users. User A has larger absolute values for $F_{ns}$ and $F_{pn}$ than $F_{sp}$, which also bears out the claim that user A's abnormal neighborhood has led to SATAR's decision.
\end{itemize}

The case study in Figure ~\ref{fig:case} demonstrates that SATAR identifies bot users by jointly evaluating their semantic, property and neighborhood information. Affinity values of our proposed Co-Influence aggregator provides explanation to SATAR's decisions.

\section{Conclusion and Future Work}
\label{sec:conclusion}
Social media bot detection is attracting growing attention. We proposed SATAR, a self-supervised approach to Twitter account representation learning and applied it to the task of bot detection. SATAR aims to tackle the challenges of generalizing in real-world scenarios and adapting to bot evolution, where previous efforts failed. We conducted extensive experiments to demonstrate the efficacy of SATAR-based bot detection in comparison to competitive baselines. Further exploration proved that SATAR also succeeded in generalizing on the real Twittersphere and adapting to different generations of Twitter bots. In the future, we plan to apply the SATAR representation learning framework to other tasks in the social media domain such as fake news detection and content recommendation.

\bibliographystyle{ACM-Reference-Format}
\bibliography{sample-base}

\end{document}